\newif\ifsubmode
\submodefalse

\newif\ifprintfig
\printfigfalse

\ifsubmode
  \documentstyle[12pt,aasms4,epsf]{article}
  \received{}
  \accepted{}
  \journalid{}{}
  \articleid{}{}
\else
  \documentstyle[11pt,aaspp4,epsf]{article}
  \slugcomment{{\it accepted for publication in The Astrophysical Journal
Supplements}}
\fi

\lefthead{Cretton et al.}
\righthead{Axisymmetric Three-Integral Models for Galaxies}

\def\etal{{et al.~}}

\def\Msun{{\rm\,M_\odot}}

\def\kms{{\rm\,km\,s^{-1}}}
\def\pc{{\rm\,pc}}

\def\Mpc{{\rm\,Mpc}}

\def\onetwo{{\textstyle {1 \over 2} \displaystyle}}

\def\DF{{\rm DF}}
\def\VP{{\rm VP}}
\def\RMS{{\rm RMS}}

\def\xp{x'}
\def\yp{y'}
\def\zp{z'}
\def\rp{r'}
\def\thetap{\theta'}
\def\qp{q'}
\def\Rdot{{\dot R}}
\def\zdot{{\dot z}}
\def\phidot{{\dot \phi}}
\def\Rzvc{R_{\rm zvc}}

\def\zzvc{z_{\rm zvc}}
\def\vlos{v_{\rm los}}

\def\vrsq{\langle v_r^2 \rangle}

\def\vthsq{\langle v_{\theta}^2 \rangle}
\def\vphisq{\langle v_{\phi}^2 \rangle}

\def\dint{\int\!\!\!\int}
\def\trint{\int\!\!\!\int\!\!\!\int}

\def\d{{\rm d}}

\def\fr#1#2{{\textstyle {#1 \over #2}}}

\def\spose#1{\hbox to 0pt{#1\hss}}
\def\lta{\mathrel{\spose{\lower 3pt\hbox{$\sim$}}
    \raise 2.0pt\hbox{$<$}}}
\def\gta{\mathrel{\spose{\lower 3pt\hbox{$\sim$}}
    \raise 2.0pt\hbox{$>$}}}

\begin{document}

\title{Axisymmetric Three-Integral Models for Galaxies}

\author{N.~Cretton, P.~Tim de Zeeuw,}
\affil{Sterrewacht Leiden, Postbus 9513, 2300 RA Leiden, The Netherlands}

\author{Roeland P.~van der Marel,}
\affil{Space Telescope Science Institute, 3700 San Martin Drive, Baltimore, 
	MD 21218}

\author{Hans--Walter Rix\altaffilmark{1,2}}
\affil{Steward Observatory, University of Arizona, Tucson,
       AZ 85721}


\altaffiltext{1}{Alfred P.~Sloan Fellow}

\altaffiltext{2}{present address: Max-Planck-Institut f\"ur Astronomie, 
Koenigstuhl 17, Heidelberg, Germany}


\ifsubmode\else
\clearpage\fi


\ifsubmode\else
\baselineskip=14pt
\fi


\begin{abstract}
We describe an improved, practical method for constructing galaxy models
that match an arbitrary set of observational constraints, without
prior assumptions about the phase-space distribution function (DF).
Our method is an extension of Schwarzschild's orbit superposition
technique. As in Schwarzschild's original implementation, we compute a
representative library of orbits in a given potential.  We then
project each orbit onto the space of observables, consisting of
position on the sky and line-of-sight velocity, while properly taking
into account seeing convolution and pixel binning. We find the
combination of orbits that produces a dynamical model that best fits
the observed photometry and kinematics of the galaxy. A new
element of this work is the ability to predict and match to the data
the full line-of-sight velocity profile shapes. A dark component (such
as a black hole and/or a dark halo) can easily be included in the
models.

In an earlier paper (Rix et al.) we described the basic principles,
and implemented them for the simplest case of spherical geometry. Here
we focus on the axisymmetric case.  We first show how to build galaxy
models from individual orbits. This provides a method to build models
with fully general DFs, without the need for analytic integrals of
motion. We then discuss a set of alternative building blocks, the
two-integral and the isotropic components, for which the observable
properties can be computed analytically. Models built entirely from
the two-integral components yield DFs of the form $f(E,L_z)$, which
depend only on the energy $E$ and angular momentum $L_z$. This
provides a new method to construct such models. The smoothness of the
two-integral and isotropic components also makes them convenient to
use in conjunction with the regular orbits.

We have tested our method, by using it to reconstruct the properties
of a two-integral model built with independent software.  The test
model is reproduced satisfactorily, either with the regular orbits, or
with the two-integral components.  This paper mainly deals with the
technical aspects of the method, while applications to the galaxies
M32 and NGC 4342 are described elsewhere (van der Marel et al.,
Cretton \& van den Bosch).
\end{abstract}




\clearpage


\section{Introduction}
\label{s:intro}

In order to understand the structure and dynamics of a galaxy, one
needs to measure the total gravitational potential as well as the
phase-space distribution function (DF) of the constituent stars. The
DF specifies the distribution of the stars over position and velocity,
and hence provides a full description of the galaxy. For a particular
galaxy, one needs to explore which combinations of potential and DF
are consistent with the available observations (surface brightness and
kinematics). Several methods have been devised to tackle this problem.

The direct calculation of the DF generally requires analytic knowledge
of the integrals of motion, and has been restricted in the past to a
number of special cases: (i) spherical or other integrable potentials
(e.g., Dejonghe 1984, 1986; Bishop 1987; Dejonghe \& de Zeeuw 1988;
Gerhard 1991; Hunter \& de Zeeuw 1992); (ii) nearly integrable systems
where perturbation theory can be applied (Saaf 1968; Dehnen \& Gerhard
1993); or (iii) the subset of axisymmetric models in which the DF is
{\it assumed} to depend only on $E$ and $L_z$ (Hunter \& Qian 1993;
Dehnen \& Gerhard 1994; Kuijken 1995; Qian \etal 1995, hereafter Q95;
Magorrian 1995; Merritt 1996b, hereafter M96b). Numerical calculations
of orbits in axisymmetric potentials have shown that most of the
orbits admit a third integral, which in general is not known
analytically (e.g., Ollongren 1962). There is no a priori physical
reason to expect the DF to depend only on the two classical integrals,
and in fact, there are indications for both elliptical galaxies
(Binney, Davies \& Illingworth 1990) and halos of spirals (Morrison,
Flynn \& Freeman 1990) that the DF must depend also on the third
integral. For the solar neighborhood it has been known for a long time
that there must be such a dependence (e.g., Binney \& Merrifield
1998).

Schwarzschild (1979, 1982) devised an elegant method to circumvent our
ignorance of analytic integrals of motion and to build numerically
self-consistent equilibrium models of galaxies. Richstone (1980, 1984)
used this technique to construct axisymmetric scale-free models. It
was applied to a variety of models (spherical, axisymmetric and
triaxial) by Richstone and collaborators (see e.g., Richstone \&
Tremaine 1984, 1985; Levison \& Richstone 1985, 1987; Katz \&
Richstone 1985). Pfenniger (1984) used Schwarzschild's method to build
two-dimensional models of barred galaxies and Merritt \& Fridman
(1996) and Merritt (1996a) used it to build a number of triaxial
models with cusps. Zhao (1996b) modeled the Galactic bar using similar
techniques.  Schwarzschild's original experiment reproduced
self-consistently a triaxial mass distribution, but as shown by
Pfenniger (1984), one can easily include kinematic constraints in the
models. Levison \& Richstone (1985) modeled the observed mean
line-of-sight velocities $V$ and velocity dispersions $\sigma$ to
estimate the amount of counter-rotation in some well-observed
galaxies.

Recent advances in detector technology have made it possible to
measure full line-of-sight velocity profile (VP) shapes, instead of
only the first two moments $V$ and $\sigma$ (e.g., Franx
\& Illingworth 1988; Rix \& White 1992; van der Marel \& Franx 1993, 
hereafter vdMF; Kuijken \& Merrifield 1993). This provides further
constraints on the dynamical structure of galaxies. Rix \etal (1997,
hereafter R97) took advantage of this development, and extended
Schwarzschild's scheme to model VP shapes. They applied it to
spherical models for the E0 galaxy NGC~2434, and showed that the
observations imply the presence of a dark halo.  Here we consider
axisymmetric models and show how to use the extended Schwarzschild
method to construct fully general three-integral models that can match
any set of kinematic constraints.  Independent implementations of the
software were written by N.C. and R.v.d.M. A summary of this
development is given by de Zeeuw (1997).  In an earlier paper (van der
Marel \etal 1998, hereafter vdM98; see also van der Marel \etal 1997)
we applied this modeling technique to the compact E3 elliptical M32,
for which previous modeling had suggested the presence of a central
massive black hole (BH) (e.g., Q95; Dehnen 1995). Cretton \& van den
Bosch (1999) describe an application to the edge-on S0 galaxy
NGC~4342. Other groups are in the process of developing similar
techniques to the one described here (e.g., Richstone \etal 1997; see
also: Emsellem, Dejonghe \& Bacon 1999; Matthias \& Gerhard 1999).

This paper is organized as follows. In Section~\ref{s:models} we
describe step by step how to construct the models (see
Figure~\ref{f:flowchart}). We first discuss the mass models that we
consider (Section~\ref{ss:massmodel}). We describe how we choose a
grid in integral space that yields a representative library of orbits
(Section~\ref{ss:orbitschoice}), how these orbits are calculated
numerically (Section~\ref{ss:orbitcalc}), how their properties are
stored on a number of grids (Section~\ref{ss:properties}), and how we
model all aspects of the data taking and analysis, such as seeing
convolution, pixel binning, and extraction of VPs
(Section~\ref{ss:modelobs}). We then present the method that we employ
to determine the non-negative weight of each orbit (i.e., the number
of stars traveling on each orbit), such that the global superposition
of orbits produces a consistent model that best fits the observations
(Section~\ref{ss:constraintsfit}). Lastly, we discuss how we include
optional smoothness constraints in the models
(Section~\ref{ss:regul}).  In Section~\ref{s:components} we describe a
set of alternative building blocks, the two-integral and isotropic
components, for which the observable properties can be computed
analytically. The smoothness of these components makes them a
convenient tool to use in conjunction with the regular orbits
described in Section~\ref{s:models}. Models can also be built entirely
of these components, to obtain models with DFs of the form $f(E,L_z)$
or $f(E)$. In Section~\ref{s:tests} we describe the tests
that we have performed to establish the accuracy of our method. We
present our conclusions in Section~\ref{s:concl}.

\section{Construction of Dynamical Models} 
\label{s:models}

\subsection{Mass Model}
\label{ss:massmodel}

We study dynamical models in which all relevant quantities are
axisymmetric, and symmetric with respect to the equatorial plane
$z=0$. 
%
%
It is sufficient to have the total gravitational potential, $\Phi
\equiv \Phi_{\star} + \Phi_{\rm dark}$, and the forces, ${\vec \nabla} \Phi$
available and tabulated on a grid, such that their values at any point
can be recovered through interpolation. This is important, because the
structure of real galaxies can be very complicated, and is not always
well described in terms of analytical functions.

While the method works for arbitrary radial density profiles, it
proves convenient for the purpose of presenting and testing our
technique to consider models in which the mass density of the luminous
material, $\rho_{\star}$, does have an analytical form:
\begin{equation}
  \rho_{\star} (R,z) = \rho (s) = 
  \rho_0 \Bigl [ {s \over b} \Bigr ]^{\alpha} 
      \Bigl (1 + \Bigl [ {s \over b} \Bigr ]^{\gamma} \Bigr )^{\beta} 
      \Bigl (1 + \Bigl [ {s \over c} \Bigr ]^{\epsilon} \Bigr )^{\delta},
\label{densitylaw}
\end{equation}
where $s$ is defined as $s^2 = R^2 + (z/q)^2$. This is an axisymmetric
generalization of the spherical models studied previously by, e.g.,
Dehnen (1993), Tremaine \etal (1994) and Zhao (1996a), and includes
the $(\alpha,\beta)$ models of Q95 as a limiting case. The model has a
constant axial ratio $q$ that does not vary with radius. The
parameters $b$ and $c$ are characteristic lengths. At small radii ($r
\ll b$) the density has a central cusp with logarithmic slope $\alpha$
(when $\alpha<0$). At intermediate radii ($b \ll r \ll c$) the density
falls off as $\rho_{\star} \propto r^{\alpha + \gamma\beta}$, while at
large radii ($r \gg c$) $\rho_{\star} \propto r^{\alpha +
\gamma\beta + \epsilon\delta}$. When viewed at an inclination angle
$i$, the isophotes are ellipses of axial ratio $\qp = (\cos^2 i + q^2
\sin^2 i)^{1/2}$. The luminosity density is
$j=\rho_{\star}/\Upsilon$, where $\Upsilon$ is the average
mass-to-light-ratio of the luminous material, which we assume to be
constant.

For these models, the gravitational potential and the associated
radial and vertical forces can all be obtained from one-dimensional
(usually numerical) integrals (cf.~eqs.~[2.10]--[2.12] of Q95).  We
calculate the potential and forces in this way and tabulate them on a
fine polar $(r,\theta)$ grid, with logarithmic sampling in radius and
linear sampling in the angle. These tabulated values are used for the
subsequent orbit calculations. It is straightforward to add the
contributions from a dark component to the potential and the forces,
as required for models with, e.g., a BH or a dark halo. In the
case of a BH these contributions need not be tabulated,
because they are known analytically.

For density distributions that are not stratified on similar
concentric spheroids one must use more general techniques to calculate
the gravitational potential and the associated forces. One possibility
to determine these, while at the same time fitting a complicated
surface brightness distribution, is to use a Multi-Gaussian Expansion
(Emsellem \etal 1994). We do this in our modeling of the S0 galaxy NGC
4342 (Cretton \& van~den~Bosch 1999). Another possibility is to obtain
$\rho_{\star}$ through non-parametric deprojection of an observed
surface brightness distribution (e.g., Dehnen 1995), and calculate the
potential from a multipole or other expansion (e.g., Hernquist \&
Ostriker 1992; Zhao 1996a).

\subsection{Choice of Orbits}
\label{ss:orbitschoice}

The results obtained with our modeling technique should not depend on
the details of the orbit library. To achieve this, the library must
represent the full variety of orbits in the given potential. In this
section we describe how we have chosen to select orbits in order to
fulfill this requirement.

In axisymmetric models, all orbits conserve at least two isolating
integrals of motion: the energy $E$ and the vertical component of the
angular momentum $L_z$. Numerical studies have shown that many orbits
conserve an additional third isolating integral $I_3$, which is
usually not known analytically (see e.g., Ollongren 1962; Innanen \&
Papp 1977; Richstone 1982; Dehnen \& Gerhard 1993).
%
%
These {\em regular} orbits are specified
completely by the integrals of motion, and can be labeled by the
values of $E$, $L_z$ and $I_3$.

For each energy $E$, there is one circular orbit in the equatorial
plane, which has radius $R_c$ and velocity $V_c^2 = R_c ({\partial \Phi
/ \partial R})_{(R_c,0)}$. The angular momentum of this orbit, $R_c
V_c$, is the maximum angular momentum at the given energy: $L_{\rm
max}(E)$. We sample the energies in the model by adopting a
logarithmic grid in $R_c$. Each $R_c$ defines an energy $E$ through
the implicit relation $E = \Phi(R_c,0) + {1\over2}V_c^2$. The orbits
in the model have $R_c \in [0,\infty)$. However, it is sufficient to
adopt a grid of $N_E$ values that covers only a finite range,
$R_{c,{\rm min}}$ to $R_{c,{\rm max}}$, chosen so as to contain all
but a negligible fraction of the total mass of the system. At each
energy we sample the range of possible $L_z$ values by adopting a grid
in the quantity $\eta \equiv L_z / L_{\rm max}(E)$ ($\eta \in
[-1,1]$). Orbits with both $L_z > 0$ and $L_z < 0$ are included in the
library, but the $L_z < 0$ orbits need not be calculated; they are
obtained from the $L_z > 0$ orbits by reversing the velocity vector at
each point along the orbit. We have calculated orbits for $N_{\eta}$
values of $\eta$, spaced linearly between $\epsilon_1$ and
$1-\epsilon_1$, where $\epsilon_1$ is a small number. For numerical
reasons, the special values $\eta = 0$ (radial orbits) and
$\eta = 1$ (the circular orbit in the equatorial plane) are presumed
to be represented by their closest neighbors on the grid, but are not
included explicitly.

In an axisymmetric potential the orbit reduces to a two-dimensional
motion in the meridional $(R,z)$ plane in an effective gravitational
potential $\Phi_{\rm eff} = \Phi + {1\over2} L_z^2/R^2$ (e.g., Binney
\& Tremaine 1987, hereafter BT). For fixed $(E,L_z)$, the position of
a star is restricted to the region bounded by the `zero-velocity
curve' (ZVC), defined as the curve of values $(R,z)$ that satisfy the
equation $E = \Phi_{\rm eff}$, and hence $v_R = v_z = 0$.
Figure~\ref{f:fig_ZVCs} illustrates ZVCs in the meridional plane.  A
regular orbit admits a third isolating integral, $I_3$, that restricts
its motion to a sub-region of the full region of phase-space
accessible at the given $(E,L_z)$. This is illustrated in
Figure~\ref{f:anorbit}, which shows a regular orbit viewed in the
meridional plane. In our method we have chosen a numerical
representation of $I_3$ that can be used to label the orbit. Every
orbit with $L_z \not=0$ touches the ZVC (Ollongren 1962). As suggested
by Levison \& Richstone (1985), we take the $R$ coordinate of the
`turning point' on the ZVC (i.e., the intersection of the orbit with
the ZVC), denoted by $\Rzvc$, as the third parameter to specify the
orbit. At every $(E,L_z)$ there is exactly one orbit that touches the
ZVC at only one value, $R_{\rm thin}$, of $R$: the so-called `thin
tube' orbit (see Figure~\ref{f:anorbit}).  All other regular orbits
touch the ZVC for at least two values of $\Rzvc$, one smaller than
$R_{\rm thin}$ and one larger than $R_{\rm thin}$. To sample the
orbits at a given $(E,L_z)$, we calculate trajectories that are
started with $v_R = v_z = 0$ from the ZVC, at a given radius $\Rzvc$
(this radius determines $v_{\phi}$ according to $v_{\phi} =
L_z/R$). Not every orbit launched in this way necessarily admits a
third integral, since irregular orbits also touch the ZVC. Our orbit
library therefore includes both regular and irregular orbits, and, as
we shall see in Section ~\ref{ss:orbitcalc}, we have found it
unnecessary to distinguish between them in all our tests and
applications to date.  To reduce redundancy in the library it is
sufficient to consider only orbits with $\Rzvc \in [R_{\rm
thin},R_{\rm max}]$, where $R_{\rm max}$ is the radius at which 
the ZVC intersects the plane $z=0$.

For the orbit library we have chosen to use $N_{I_3}$ values of
$\Rzvc$. Each point $(\Rzvc,\zzvc)$ on the ZVC is determined with the
help of an angle $w$, which is sampled linearly between 0 and $w_{\rm
thin}$ (see Figure~\ref{f:anorbit}). For numerical reasons, the
special values $\Rzvc = R_{\rm thin}$ (thin tube orbit) and $\Rzvc =
R_{\rm max}$ (equatorial orbit) are presumed to be represented by
their respective closest neighbor on the grid, but are not included
explicitly. Finding the starting point for the periodic orbit, 
$R_{\rm thin}$, is straightforward (see e.g. Pfenniger \& Friedli, 1993).

It is sufficient to calculate only orbits that are started from the
ZVC with $z>0$. Orbits started with $z<0$ are obtained from those
started with $z>0$ by reversing the sign of $z$ and $v_z$ at each
point along the orbit. Most orbits are themselves symmetric with
respect to the equatorial plane (see e.g., Figure~\ref{f:anorbit}), so
that this operation is redundant. However, this is not true for, e.g.,
the orbits parented by the 1/1 resonance between the $R$ and
$z$-motion (see Figure~\ref{f:orbits_bh} below, or Figure~8 of Richstone
1982). Since we are only interested in constructing models that are
symmetric about the equatorial plane, we do not view the orbits
started with $z > 0$ and $z < 0$ from the ZVC as separate building
blocks, but instead we consider only their sum.

The grid in $(R_c,\eta,\Rzvc)$ completely specifies the orbit
library. Appropriate choices for the parameters that characterize this
grid are discussed in Section~\ref{s:tests}.

\subsection{Orbit Calculation}
\label{ss:orbitcalc}

For each $(R_c,\eta,\Rzvc)$ we calculate a trajectory, started from
the ZVC as described in Section~\ref{ss:orbitschoice}. We have used
several standard integration algorithms, including the Bulirsch-Stoer
integrator (Press \etal 1992) and the Runge-Kutta-Fehlberg algorithm
(Fehlberg 1968). We have experimented with both and found equivalent
results. The former algorithm was used in vdM98. Here we use the
Runge-Kutta-Fehlberg algorithm.

The results of the orbit calculations were used to approximate the
`orbital phase-space density' for each trajectory. Each phase-point
along a calculated orbit was assigned a weight equal to the time step
at that point, divided by the total integration time.

This procedure results in density distributions in phase-space,
$\DF_{\rm traj}$ and its corresponding spatial density $\rho_{\rm
traj}$.  Orbits were calculated in the meridional plane, but all six
phase-space coordinates are needed. The azimuthal velocity
$v_{\phi}=L_z/R$ is completely specified.  However, for projection
onto the sky, the azimuthal angle $\phi \in [0,2\pi]$ is also
required. The distribution of stars over $\phi$ is homogeneous,
because of the assumed symmetry. The weight at each time step was
therefore divided into a number of equal `sub-weights', and each was
assigned a random $\phi$. Furthermore, each sub-weight was divided in
two, and one of the two parts was assigned phase-coordinates with
$(z,v_z)$ multiplied by $-1$. This corrects for the fact that only
orbits started with $z>0$ from the ZVC were calculated
(cf.~Section~\ref{ss:orbitschoice}). The trajectories should be
integrated long enough so that the orbital phase-space densities no
longer change significantly with time. Pfenniger (1984) proposed to
check directly for the convergence of the orbital mass
distribution. However, this may take a very long time, especially for
orbits that are unusually close to a high-order resonance, for orbits
at very large radii or at very small radii close to a BH, and for
irregular orbits (see also Merritt \& Fridman 1996). We have used a
cruder approach, in which we calculated each orbit for a fixed number
($\sim 200$) of characteristic orbital periods. We found this to be
sufficient for our purpose; longer integrations yield final models
that are not significantly different. This is because the `noise' in
our modeling is dominated by the representation of phase space through
a coarse discrete grid.

Orbits can have sharp edges in both the spatial and velocity
dimensions. We found that a simple scheme to obtain smoother densities
yielded slightly more accurate results for the final orbit
superposition. To take into account the fact that each energy $E$ in
the orbit catalog represents all energies in some bin $[E_1,E_2]$
around it (defined by the choice of energy grid), a random energy
${\tilde E}$ was drawn from the range $[E_1,E_2]$ for each normalized
time step. The corresponding phase-coordinates $({\vec r,\vec v})$
were then translated to the energy ${\tilde E}$, by replacing them by
$([{\tilde R}_c/R_c]{\vec r}, [{\tilde V}_c/ V_c]{\vec v})$, where
$R_c$, ${\tilde R}_c$ and $V_c$, ${\tilde V}_c$ are the radii and
circular velocities of the circular orbits at the energies $E$ and
${\tilde E}$, respectively. This ``dithering'' approach is only
approximately correct (it assumes that the potential is locally
scale-free), but was found to work well in practice.

\subsection{Storing the Orbital Properties}
\label{ss:properties}

For each orbit we store both the intrinsic properties and the
projected properties. The intrinsic properties are necessary to test
for consistency of the final model. We store $\rho_{\rm traj}$ on an
$(r,\theta)$ grid in the meridional plane, logarithmic in $r$ and
linear in $\theta \in [0,{\pi\over2}]$. Angles $\theta > {\pi\over2}$
need not be stored separately, because of symmetry with respect to the
equatorial plane. We also store the lowest-order velocity moments of
each orbit ($\rho_{\rm traj} \langle v_i \rangle, \rho_{\rm traj}
\langle v_i v_j \rangle, i,j=r,\theta,\phi$) on the same grid, so 
as to be able to study the intrinsic dynamical structure of the final model.

The projected properties are necessary for comparison to observable
quantities, such as the projected surface brightness and line-of-sight
VP shapes. Only three coordinates of phase-space are available for
comparison with observations: the projected positions $\xp$, $\yp$
(which we choose to be aligned with the photometric major and minor
axis), and the line-of-sight velocity, $\vlos (\equiv
v_{\zp})$. Given an inclination angle $i$ of the galaxy
($i=90^{\circ}$ means edge-on), these are related to the usual
cylindrical coordinates $(R, z, \phi)$ in the following way:
\begin{eqnarray}
  \xp   & = & R \sin\phi                          , \nonumber \\
  \yp   & = & -R \cos i \cos \phi + z \sin i      , \nonumber \\
  \vlos & = & (v_R \cos\phi - v_\phi \sin\phi) \sin i + v_z \cos i . 
\label{projectiontransform}
\end{eqnarray}
To have the projected properties of the orbits accessible, we store
their phase-space densities both on an $(\rp,\thetap)$ grid on the
projected plane of the sky (with similar properties as the intrinsic
$(r,\theta)$ grid), and on a Cartesian $(\xp,\yp,\vlos)$ data cube
(see Section~\ref{ss:modelobs}). The former is used to reconstruct the
projected surface density of the model. The latter is used to model
observed kinematical quantities. The spatial grid size $(\Delta
x,\Delta y)$ of the $(\xp,\yp,\vlos)$ cube is chosen to provide 2--5
times higher spatial resolution than the pixel size of the available
kinematical observations. If observations with very different
resolution are available for a galaxy (e.g., very high spatial
resolution HST data in the central arcsec, and lower-resolution
ground-based data out to an effective radius), it is best to store the
data on two or more cubes with different spatial grid sizes and
extents. During the orbit calculations we then store the phase-space
densities simultaneously on all $(\xp,\yp,\vlos)$ cubes. Only the $\xp
\geq 0$ half of each cube needs to be stored, because each orbit has
the same weight at $(\xp,\yp,v_{\rm los})$ as at
$(-\xp,-\yp,-\vlos)$. The size $\Delta v$ of the velocity bins on the
($\xp, \yp, \vlos$) cube(s) must be chosen to provide a proper
sampling of the observed VPs. In practice we use $\sim 50$--100 bins
between $(-N_{\sigma}
\sigma_{\rm max}, N_{\sigma} \sigma_{\rm max})$, where 
$\sigma_{\rm max}$ is the largest observed dispersion, and 
$N_{\sigma} = 4$--8.

\subsection{Modeling Observed Kinematical Quantities}
\label{ss:modelobs}

Point-spread-function (PSF) convolution is essential when comparing
model quantities with observed kinematical quantities in the central
regions of galaxies. Seeing convolution correlates information in the
two spatial dimensions $\xp,\yp$, but not in $\vlos$:
\begin{equation}
  F_{\rm conv} (\xp_0, \yp_0, \vlos) = F \otimes {\rm PSF} =
    \dint F(\xp,\yp,\vlos) \> {\rm PSF} (\xp-\xp_0,\yp-\yp_0) \>
    {\d}\xp \> {\d}\yp ,
\label{convoldef}
\end{equation}
where $F$ is the function to convolve, PSF is the point-spread
function, and $F_{\rm conv}$ is the result of the convolution of $F$
with the PSF. The final model is a linear superposition of the orbits,
so the $(\xp,\yp,\vlos)$ cubes for each orbit may be individually
convolved with the PSF. As in R97, we do the convolution for each
velocity slice efficiently by multiplications in Fourier space, using
Fast Fourier Transforms (e.g., Press \etal 1992).

Kinematical data is generally obtained either through small, discrete
apertures, along a number of slit-positions, or may derive from
two-dimensional integral field spectroscopy (e.g., Bacon \etal 1995). Any
setup of this kind can be simulated by our models, including possible
spatial binning along a slit. For each observational `aperture', we
choose the $(\xp,\yp,\vlos)$ cube with the most appropriate cell
size, convolve it with the relevant PSF, and bin the results spatially
over the aperture size. This yields a one-dimensional velocity
histogram, for each orbit and for each observation. Examples of such
`orbital VPs' are shown in Figure~\ref{f:orbitalVPs}.

Kinematical observations provide information on the line-of-sight VPs
\begin{equation}
  \VP (\xp, \yp, \vlos) = \trint
              \DF \> {\d} v_{\xp} \> {\d} v_{\yp} \> {\d} \zp   ,  
\label{VPequ}
\end{equation}
at different positions $(\xp,\yp)$ on the projected face of a galaxy.
In practice, the normalization of $\VP (\xp, \yp, \vlos)$ is based on
the photometric data. It is often useful to parametrize observed VPs
by a few numbers. A common choice for such a VP parametrization is the
Gauss-Hermite expansion. We follow the notation of vdMF, in which the
VP is represented as
\begin{equation}
  \VP (\vlos) = {\alpha(w) \over \sigma} \sum_{l=0}^{N} h_l H_l(w),
\label{VPdefvdMf}
\end{equation}     
with
\begin{equation}
  w = (\vlos - V)/\sigma ,\quad 
  \alpha (w) = {1 \over {\sqrt{2\pi}}} \> {\rm e}^{-w^2/2} .
\label{defvdMf}
\end{equation} 
The $h_l$ are the Gauss-Hermite moments (hereafter GH-moments)
defined by
\begin{equation}
  h_l = {2\sqrt{\pi}} \int_{-\infty}^{\infty} \VP (\vlos) \>
    \alpha (w) \> H_l(w) \> \d \vlos \qquad  (l=0,\ldots,L) .   
\label{hl}
\end{equation}
Each $H_l$ is a Hermite polynomial (see Appendix~A of vdMF). The
quantities $V$ and $\sigma$ characterize the `weighting function',
$\alpha(w) H_l$, in the integral (\ref{hl}). When describing
observations, $V$ and $\sigma$ are usually taken to be the velocity
and dispersion of the Gaussian that best fits the observed VP. With
this choice, $h_1 \equiv h_2 \equiv 0$. 
GH-moments of higher order describe deviations from a Gaussian.
Only the moments of order $L \leq 6$ are generally measured from
galaxy spectra, due to the finite spectral (and thus velocity)
resolution of the observations.

If we envision galaxies as consisting of orbital building blocks, then
the overall VP measured for a given observation is just the
superposition of the individual orbital VPs. Similarly, the observed
GH-moments are just a linear superposition of the GH-moments of the
individual orbital VPs, provided that the observed $V$ and $\sigma$
for the given observation are used in the weighting function
$\alpha(w) H_l (w)$. Thus, as described in detail in R97, to fit the
kinematical observations we may restrict ourselves to solving a linear
superposition problem for the Gauss-Hermite moments.  The constraints
are then that $h_1 = h_2 = 0$, and the $h_l$ with $l \geq 3$ should
equal their observed values. It must be stressed that this approach is
general, and assumes neither that the observed VPs are well-described
by the lowest-order terms of a GH-series, nor that the orbital VPs are
well-described by the lowest-order terms of a GH-series. Nonetheless,
if a full non-parametric estimate of the observed VPs is available
there is no need to restrict the analysis to the lowest-order
GH-moments. Our technique can just as easily fit the individual
velocity bins of the observed VPs.

\subsection{Fitting the Constraints}
\label{ss:constraintsfit}

Constructing a model consists of finding a weighted superposition of
the stellar orbits in the library that reproduces two sets of
constraints:

\begin{itemize}
\item {\bf Consistency constraints} for the stellar luminosity distribution. 
The model should reproduce the initially assumed luminous stellar
density $\rho_{\star}(R,z)$ (Section~\ref{ss:massmodel}), for each
cell of the meridional $(r,\theta)$ grid, for each cell of the
projected plane $(\rp,\thetap)$ grid, and for each aperture for which
there is kinematical data. In theory, it is sufficient to fit only the
meridional plane masses, because projected densities are then fit
automatically. In practice this might not exactly be the case, because
of discretization. To circumvent this, the projected masses may be
included as separate constraints. Note that for axisymmetric models
the projected density does not uniquely specify the intrinsic density
(e.g., Rybicki 1986; Gerhard \& Binney 1996).

\item {\bf Kinematical constraints}. The model should reproduce the observed
kinematics of the galaxy, including VP shapes. As discussed, we
express this as a set of linear constraints on the GH-moments of the
VPs (R97).
\end{itemize}

Finding the orbit superposition that best fits these constraints
amounts to solving a linear problem, which can be written in matrix
notation as ${\bf B} {\vec \gamma} = {\vec c}$ (R97). The matrix {\bf
B} contains the mass that each orbit contributes to each relevant
intrinsic or projected grid cell, and the GH-moments that each orbit
contributes to each kinematical observation. The vector ${\vec c}$
contains the mass predicted by $\rho_{\star}(R,z)$ for each relevant
intrinsic or projected grid cell, and the observed GH-moments for all
kinematical observations; the vector ${\vec \gamma}$, which is to be
solved for, contains the weight of each orbit, i.e., the total mass of
stars on each orbit. These weights should be non-negative.
Using the terminology introduced in Section~\ref{ss:modelobs}, the
 basic Schwarzschild equation becomes 
\begin{equation}
  \sum_{j=1}^{N_{\rm orbits}} \gamma_j \VP_{ij} = \VP_i ,
\label{basic_Sch_equation}
\end{equation} 
with $\VP_{ij}$ the individual VP of orbit $j$ at constraint point
$i$, and $\VP_{i}$ the observed VP for the same constraint point.

The superposition problem can be expressed as a non-negative least
squares (NNLS) fit for the above matrix equation. We have used the
NNLS routine of Lawson \& Hanson (1974) to solve it. The NNLS routine
finds a combination of non-negative orbital occupancies (which need
not be unique) that minimizes the usual ${\cal L}^2$ norm $|| \, {\bf
B} {\vec \gamma} - {\vec c} \, ||$. This norm can be viewed as a
$\chi^2$ quantity that measures the quality of the fit to the
constraints. The NNLS routine always finds a best solution. It need
not be acceptable in light of the observations; this must be assessed
through the $\chi^2$ of the best fit, and by comparison of the model
predictions to the constraints.

As is customary in least-squares fitting, the model predictions for
each constraint and the actual constraint values (the elements of the
vector ${\vec c}$) are weighted by the errors in the
constraints. Observational errors are available for the kinematical
constraints. In principle one would like the consistency constraints
to be fit with machine precision. It turns out that this is generally
unfeasible, because of discretization. It was found that models with
no kinematical constraints could at best simultaneously fit both the
intrinsic and the projected masses with a fractional error of $\sim 5
\times 10^{-3}$ (when using $\sim 1000$ orbits). We therefore assigned
fractional errors of this size to the masses in the consistency
constraints.  In principle one would like to include also the
observational surface brightness errors in the
analysis. Unfortunately, this requires the exploration of a large set
of three-dimensional mass densities (that all fit the surface
photometry to within the errors), which is prohibitively
time-consuming.

\subsection{Regularization}
\label{ss:regul}

Our orbit superposition models are not generally smooth in integral
space, as a result of the `ill-conditioned' numerical nature of the
NNLS matrix equation that is being solved. There are no physical
theories that describe exactly how smooth the DF of a stellar system
should be, but some degree of smoothness should be expected. Our
technique can be extended in a straightforward manner to yield smooth
solutions, by adding linear regularization constraints to the NNLS
matrix equation (e.g., Press \etal 1992; Merritt 1993).  This has the
same effect as the addition of `maximum entropy' constraints
(Richstone \& Tremaine 1988). For linear regularization, each
regularization constraint must be of the form
\begin{equation}
  \sum_{k} s_{k,l} \, \gamma_k = 0 \pm \Delta    ,
\end{equation}
thus providing an extra row to the matrix equation. The $\gamma_k$ are
the orbital weights that make up the vector $\vec{\gamma}$, and $l$ is
the number of the regularization constraint. The parameter $\Delta$
sets the amount of regularization.  Models with $\Delta \rightarrow
\infty$ have no regularization, while models with $\Delta \rightarrow 
0$ give infinite weight to the regularization constraints.
Alternatively, one may view this as adding a term $\lambda\, || {\bf S} 
{\vec \gamma} ||$ to the norm $|| \, {\bf B} {\vec \gamma} - {\vec c} 
\, ||$ that is minimized by the NNLS routine, where ${\bf S}$ is the 
matrix with elements $\lbrace s_{k,l} \rbrace$, and $\lambda \equiv
1/\Delta$ is a regularization parameter (Zhao 1996b).

Many choices are possible for the matrix ${\bf S}$, with the only
requirement that the norm $|| {\bf S} {\vec \gamma} ||$ should provide
a measure of the smoothness of the solution. Our choice is based on
the fact that we consider the NNLS solution $\gamma(R_c,\eta,\Rzvc)$
to be ``smooth'' if the second derivatives of the (unitless) function
$\gamma(R_c,\eta,\Rzvc)/\gamma_0(R_c)$ are small. Here the ``reference
weights'' $\gamma_0(R_c)$ are a rough approximation to the energy
dependence of the model. These are determined beforehand, e.g., by
studying the spherical isotropic limit of the given mass density. We
view the three-dimensional numerical grid in integral space as a
Cartesian lattice, and we approximate the second derivatives by second
order divided differences (eq. 18.5.10 of Press \etal 1992). We assume
that the distance between adjacent grid points on the lattice is
unity, independent of the carthesian direction in which they are
adjacent. This (arbitrarily) solves the problem that the axes of the
grid in integral space have different units and yields three
regularization constraints for each grid point $(i,j,l)$ that is not
on a boundary: $-\gamma_{i-1,j,l} + 2\gamma_{i,j,l} -\gamma_{i+1,j,l}
= 0$, $-\gamma_{i,j-1,l} + 2\gamma_{i,j,l} -\gamma_{i,j+1,l} = 0$, and
$-\gamma_{i,j,l-1} + 2\gamma_{i,j,l} -\gamma_{i,j,l+1} = 0$.

\section{Two-integral Components and Isotropic Components}
\label{s:components}

\subsection{Definition}
\label{ss:compdef}

Individual regular orbits correspond to building blocks with a DF
proportional to $\delta(E-E_0) \> \delta(L_z-L_{z,0}) \> \delta(I_3 -
I_{3,0})$, for given $(E_0,L_{z,0}, I_{3,0})$. These are not the only
building blocks that can be used to construct models. One may also use
`two-integral components', which correspond to the DF
\begin{equation}
  f^{\delta}_{[E_0,L_{z,0}]} \equiv
     C_{[E_0,L_{z,0}]} \> \delta(E-E_0) \> \delta(L_z-L_{z,0}) ,
\label{fdeltatwoint}
\end{equation}
or `isotropic components', which correspond to the DF (cf.\ Richstone
1982)
\begin{equation}
  f^{\delta}_{[E_0]} \equiv C_{[E_0]} \> \delta(E-E_0) .
\label{fdeltaE}
\end{equation}
We choose the normalization coefficients $C_{[E_0,L_{z,0}]}$ and
$C_{[E_0]}$ such that the total mass of each component is equal to
unity; explicit expressions are derived in Appendix~\ref{ss:compnorm}.

The two-integral components are smoother building blocks than the
regular orbits, since they fill completely the ZVC and do not have the
sharp edges of the regular orbits. It is useful to view them as a
particular combination of all orbits that could be numerically
integrated at the given $(E_0,L_{z,0})$, both regular orbits that fill
only a subset of the area enclosed by the ZVC (therefore admitting 3
independent integrals of motion) {\it and} irregular orbits that
occupy a larger area (admitting only 2 integrals; note that an
irregular orbit does not necessarily fill the entire phase-space
region defined by $(E_0,L_{z,0})$; see Merritt \& Valluri 1996 for a
discussion of the triaxial case). Similarly, an isotropic component is
a weighted combination of all two-integral components (i.e., all
orbits) at the given energy $E_0$. The region in space occupied by
such a component is bounded by the equipotential surface $\Phi(R,z) =
E_0$.

The two-integral and isotropic components are useful, because their
properties can be calculated semi-analytically. By using {\it only}
two-integral components in the NNLS orbit superposition, one can
construct $f(E,L_z)$ models for arbitrary spheroidal potentials. This
provides a new and convenient way of constructing such models, which
adds to the several techniques already in existence for this purpose
(Hunter \& Qian 1993; Dehnen \& Gerhard 1994; Kuijken 1995; Magorrian
1995; M96b). Using only isotropic components in the NNLS orbit
superposition is generally less useful, because these components
follow equipotential surfaces, which are rounder than isodensity
surfaces. Thus, they cannot be used to build self-consistent isotropic
axisymmetric models. R97 describe how to use them to build spherical
isotropic models. Alternatively, the two-integral and isotropic
components may be used in the superposition {\it in conjunction with}
the regular orbits. This has two advantages. First, these components
are smoother, and their inclusion therefore reduces numerical noise
that arises from the discrete representation of phase space (see also
Zhao 1996b). Second, addition of these components provides a way to
include all irregular orbits in the models.

\subsection{Velocity Profiles}
\label{ss:compVPs}

The VP of a two-integral component is obtained by substitution of the
DF of equation~(\ref{fdeltatwoint}) into equation~(\ref{VPequ}). The
resulting integral may be written as
\begin{equation}
  {\VP}_{[E_0,L_{z,0}]} (\xp,\yp,\vlos) = 
     C_{[E_0,L_{z,0}]} \int {\d}\zp \> J_{[E_0,L_{z,0}]},
\label{VPtwoint}
\end{equation}
where 
\begin{equation}
  J_{[E,L_z]} \equiv 
     \Bigl | 
        { {\partial(v_{\xp},v_{\yp})} \over {\partial(E,L_z)} }
     \Bigr |_{[E,L_z]},
\label{JEL}
\end{equation}
is the Jacobian for the change of variables from $(v_{\xp},v_{\yp})$
to $(E,L_z)$. In Appendix~\ref{ss:compjacob} we give an explicit
expression for this Jacobian. The integration in
equation~(\ref{VPtwoint}) extends over those $\zp$ for which there
exist velocities $(v_{\xp},v_{\yp})$ such that
$E(\xp,\yp,\zp,v_{\xp},v_{\yp},\vlos) = E_0$ and
$L_z(\xp,\yp,\zp,v_{\xp},v_{\yp},\vlos) = L_{z,0}$. Similarly, the VP
for an isotropic component may be written as
\begin{equation}
  {\VP}_{[E_0]} (\xp,\yp,\vlos) =
     C_{[E_0]} \int {\d}\zp \int {\d}L_z \> J_{[E_0,L_z]} .
\label{VPiso}
\end{equation}
The projected density for a two-integral or isotropic component, at
projected position $(\xp,\yp)$, is obtained as the integral of
${\VP} (\xp,\yp,\vlos)$ over $\vlos$.

Equations~(\ref{VPtwoint}) and~(\ref{VPiso}) can be used to calculate
the VPs of the two-integral and isotropic components through numerical
quadratures, without the need for calculating orbital
trajectories. The only difficulty lies in finding the domain of
integration in $\zp$. We illustrate this for the case of an edge-on
system ($i = 90^{\circ}$). In this case $J_{[E_0,L_{z,0}]} = (\zp
v_{\yp})^{-1}$, with
\begin{equation}
  v_{\yp} = \sqrt{ 2(E_0 - \Phi) - 
                     \Bigl ( {{L_{z,0} + \xp\vlos}\over{\zp}}
                     \Bigr )^2 - \vlos^2 } 
\label{vypedgeon}
\end{equation}
(cf.~eqs.~[\ref{generaljacobian},\ref{vxyproj}]). We will refer to the
expression under the square-root as $W$. The integration in
equation~(\ref{VPtwoint}) extends over those $\zp$ for which $W \geq
0$. We find the roots of $W$ numerically. We start by finding the
roots of $2 (E_0 - \Phi) - \vlos^2 = 0$. This gives an interval that
encompasses {\it all} real roots of $W$, because $2 (E_0 - \Phi) -
\vlos^2 \geq W$. Then we subdivide this interval in many ($\sim
100$--1000) small segments, and check whether the sign of $W$ differs
at the ends of each segment. If it does, we find the root between
these two points through bisection. The continuity of the resulting
VP was used to check whether all required integration domains in $\zp$
were found. For the potentials studied here, we typically find two or
four roots.

The VP calculation for edge-on {\it isotropic} components is less
complicated. The Jacobian is quadratic in $L_z$, and can be written as
$J_{[E_0,L_z]} = [(L_z^{+} - L_z)(L_z - L_z^{-})]^{-1/2}$. One can
show that $L_z^{-}$ and $L_z^{+}$ are real if $2 (E_0 - \Phi) -
\vlos^2 \geq 0$.  In this case the integral over ${\d}L_z$ in
equation~(\ref{VPiso}) extends from $L_z^{-}$ to $L_z^{+}$, and is
always equal to $\pi$.  Thus, if there is a $\zp_{\rm max}$ for which
$2 (E_0 - \Phi) - \vlos^2 = 0$, then
\begin{equation}
  {\VP}_{[E_0]} (\xp,\yp,\vlos) = 2 \pi C_{[E_0]} \zp_{\rm max} .
\label{VPisoedgeon}
\end{equation}
If there is no such $\zp_{\rm max}$ (i.e., if $\vlos$ exceeds the
escape velocity at the tangent point), then ${\VP}_{[E_0]}
(\xp,\yp,\vlos) = 0$.

Figure~\ref{f:twointVPs} shows examples of the VP of a two-integral
component along the major and minor axes of an edge-on
system. 

\section{Tests}
\label{s:tests}

\subsection{The Test Model}
\label{ss:testmodel}

The most useful tests for our axisymmetric implementation are
those for which the results can be compared to analytical results, or
to semi-analytical or numerical results that were obtained with an
independent algorithm. Models with $f(E,L_z)$ DFs have been widely
studied in the past five years (e.g., Evans 1993, 1994; Hunter \& Qian
1993; Dehnen \& Gerhard 1994; Evans \& de Zeeuw 1994; Kuijken 1995;
Q95; Magorrian 1995; M96b). Their properties can be derived
semi-analytically, and a variety of algorithms and numerical
implementations have been presented to derive the DF $f(E,L_z)$ that
generates a given luminous mass density $\rho_{\star}(R,z)$ in a given
potential $\Phi(R,z)$. These models therefore provide an ideal test
case. Here we present two tests where we use our method to
reproduce the properties of an edge-on $f(E,L_z)$ model.

We consider a model with a luminous mass density of the form
(\ref{densitylaw}), with parameters: $\alpha = -1.435$, $\beta =
-0.423$, $\gamma = \epsilon = 2.0$, $\delta = -1.298$, $b=0.55''$,
$c=102.0''$, $q=0.73$, $\rho_0 = j_0 \Upsilon M_{\odot} /
L_{\odot,V}$, $j_0 = 0.463 \times 10^5 L_{\odot,V} \pc^{-3}$ (for an
assumed distance of $0.7 \Mpc$). We calculate the potential of the
test model, $\Phi \equiv \Phi_{\star} + \Phi_{\rm dark}$, under the
assumption that $\Upsilon = 2.5$, and with the option of a nuclear
BH ($\Phi_{\rm dark} = - G M_{\rm BH} / r$) of mass $M_{\rm
BH} = 3 \times 10^6 \Msun$. All these parameters are based on the
application of our technique to the case of the galaxy M32, which was
presented in vdM98. This analogy with M32 was chosen mainly to
demonstrate the accuracy of our method for a realistic galaxy model.

The luminous mass density $\rho_{\star}(R,z)$ and potential
$\Phi(R,z)$ determine uniquely only the even part $f_{\rm
even}(E,L_z)$ of the DF, $f(E,L_z) \equiv f_{\rm even}(E,L_z) + f_{\rm
odd}(E,L_z)$. For our test model we specify (arbitrarily) the extreme
case that $f_{\rm odd}(E,L_z) = f_{\rm even}(E,L_z)$ for $L_z > 0$,
and that $f_{\rm odd}(E,L_z) = - f_{\rm even}(E,L_z)$ for $L_z < 0$
(and by definition, $f_{\rm odd}(E,0) = 0$). Thus, the $f(E,L_z)$ test
model is `maximally rotating': all the stars are rotating in the same
sense and have $L_z>0$.

First, the unique $f(E,L_z)$ DF of the test model was calculated using
the approach described in vdM98. We will refer to the resulting DF as
${\rm DF}_{\rm HQ}$ (for Hunter \& Qian, 1993). The kinematical
predictions (VPs, GH-moments, etc.) for the test model DF were
subsequently calculated using the expressions and software of Q95. The
Jeans equations were used as in Cretton \& van den Bosch (1999) to
compute the intrinsic second-order velocity moments $\vphisq$ and
$\vrsq = \vthsq$ in the meridional plane. Our tests in
Sections~\ref{ss:testtwoint} and~\ref{ss:testregorb} are aimed at
assessing how well our algorithm can reproduce the test
model properties thus calculated with independent methods. This allows
us to test all key aspects of the orbit model construction, including
the sampling of integral space, orbit calculation, discreteness
effects of the spatial grids, projection into the data cubes, seeing
convolution, and the NNLS algorithm. Hence, it is no great drawback
that our tests are restricted to two-intergal models.

\subsection{Reproducing the Test Model with Two-integral Components}
\label{ss:testtwoint}

We first describe tests of the extended Schwarzschild technique with
only two-integral components. We used an $(E,L_z)$ grid as described
in Section~\ref{ss:orbitschoice}, with $N_E = 70$, $R_{c,{\rm min}} =
10^{-4.2}$ arcsec, $R_{c,{\rm max}} = 10^{4.2}$ arcsec, $N_{\eta} =
19$, and $\epsilon_1 = 0.01$. Only components with $L_z > 0$ were
included in the superposition; the resulting models are therefore by
definition maximally rotating with a DF of the form $f(E,L_z)$. The DF
is determined uniquely by the mass density. Kinematical constraints
are therefore not required in the NNLS fit, but only constraints on
the consistency of the stellar luminosity distribution (see
Section~\ref{ss:constraintsfit}). For these, the polar $(r,\theta)$
and $(\rp,\thetap)$ grids in the meridional plane and on the projected
plane of the sky (see Section~\ref{ss:properties}) were chosen to have
16 bins in the radial coordinate between $R_{c,{\rm min}}$ and
$R_{c,{\rm max}}$, and $N_{\theta} = N_{\thetap} = 7$ bins in the
angular coordinate (a rather modest resolution, but similar tests with
finer grids yielded similar accuracies). We semi-analytically
(eq.~[4-140b] of BT) calculated the isotropic DF $f(E)$ for the
spherical version of the test model, and used the corresponding masses
on our energy grid as reference weights for the regularization (see
Section~\ref{ss:regul}).

The NNLS algorithm yields the mass on each $(E,L_z)$ grid cell, i.e.,
the integral of ${\d}M \, / \, {\d}E \, {\d}L_z$ over the grid cell.
It does {\it not} directly yield the DF $f(E,L_z)$, which by
definition is the density in the six dimensional phase-space. However,
for a two-integral model there is a simple relation between ${\d}M \,
/ \, {\d}E \, {\d}L_z$ and the DF $f(E,L_z)$, as derived in
Appendix~\ref{ss:weightstwoint}. With equation~(\ref{gammaptwoint})
the NNLS fit provides an estimate of the DF, which we will denote
${\rm DF_{NNLS}}$. We compare it with ${\rm DF_{HQ}}$ on the same
grid, but to avoid possible border effects, we restrict the comparison
to the $N_E = 50$ energy grid points with $R_{c}$ between $10^{-3}$
and $10^3$ arcsec (see Figures~\ref{f:DF_reconstruction_BH}
and~\ref{f:DF_reconstruction_no_BH} for the test models with and
without a central BH, respectively).

The DFs agree well over 10 and 20 orders of magnitude,
respectively. The inserts show the percentage errors in the DF
calculation. Note that the largest errors occur at grid points that
carry little mass, e.g., at large radii. The orbit library in these
figures is numbered as follows. For each value of $R_c(E)_j$ of the
energy, $L_z$ runs monotonically from $L_{z,{\rm min}}$ to $L_{z,{\rm
max}}$, covering $N_{\eta}$ (= 19 here) components. The next orbit
corresponds to $L_{z,{\rm min}}$ of $R_c(E)_{j+1}$, etc. This choice
of numbering causes the jagged appearance of the DF. Q95 plotted
$f(E,L_z=0)$ and $f(E,L_z = L_{z,{\rm max}})$ as a function of $E$
(see their Figure 8), and such curves would appear as envelopes in
Figures~\ref{f:DF_reconstruction_BH}
and~\ref{f:DF_reconstruction_no_BH}.

To assess the influence of smoothing on the accuracy with which our
technique recovers the DF, we have studied the dependence of the RMS
logarithmic residual
\begin{equation}
  \RMS_{\log {\rm DF}} \equiv
      \Bigl [ {1 \over {N_E N_{\eta}}}
              \sum_{i=1}^{N_E} \sum_{j=1}^{N_{\eta}}
              ( \log {\rm DF_{NNLS}} - \log {\rm DF_{HQ}})^2
      \Bigr ]^{1/2},
\label{chisqm}
\end{equation}
on the regularization parameter $\Delta$ (see Section~\ref{ss:regul}).
When $\Delta$ tends to zero, the regularization constraints receive
infinite weight. This yields a very smooth DF, but one that doesn't
fit the consistency (mass) constraints very well, and therefore
doesn't approximate $\DF_{\rm HQ}$ very well. At the other extreme,
when $\Delta$ is very large there is hardly any smoothing, and the
mathematically ill-conditioned nature of the problem yields a very
jagged solution that also doesn't match $\DF_{\rm HQ}$.
Figure~\ref{f:rms_combined} shows $\RMS_{\log \DF} (\Delta)$ for the
following three cases: (a) the case in which only the masses on the
meridional plane $(r,\theta)$ grid are included as constraints in the
NNLS fit; (b) the case in which only the masses on the projected plane
$(\rp,\thetap)$ grid are included as constraints; and (c) the case in
which both are included as constraints. In principle, deprojection of
the projected mass density is unique for an edge-on system, so these
approaches should recover equivalent results. However, this is not
true in practice because of discretization effects. In all three
cases, $\RMS_{\log {\rm DF}}$ has a minimum near ${\rm log} \Delta
\approx 2.0$, which is thus the optimum smoothing. The
value of $\RMS_{\log {\rm DF}}$ at the minimum is only mildly
different for the different cases, but (a) yields the slightly better
results. We have therefore adopted case~(a) for all our further test
calculations. The minimum $\RMS_{\log {\rm DF}}$ is $\sim 0.02$; this
corresponds to a 5\% RMS difference between ${\rm DF_{NNLS}}$ and
${\rm DF_{HQ}}$. The mass-weigthed RMS difference,
\begin{equation}
  {\rm RMS}_{\rm DF} = \Biggl [ 
    {\int\!\!\!\!\int \DF_{\rm HQ} \>
      \Bigl[{\DF_{\rm NNLS} - \DF_{\rm HQ} \over \DF_{\rm HQ}} \Bigr]^2 \>
    {\d}^3 {\vec x} \> {\d}^3 {\vec v}} \; \bigg/ \; 
    {\int\!\!\!\!\int \DF_{\rm HQ} \>
    {\d}^3 {\vec x} \> {\d}^3 {\vec v}}  \Biggr ]^{1/2}  ,
  \label{global_diff}
\end{equation}
for the model with the optimum smoothing is also $\sim 5 \%$. This
level of accuracy in the determination of the DF is similar to that
obtained with other techniques (see Gerhard \etal 1998; Matthias \&
Gerhard 1999).

Figure~\ref{f:2nd_moments_no_BH} compares the predictions for the
meridional plane velocity moments to the results of the Jeans
equations, for the model without a BH and with the optimum smoothing.
The top and bottom row of the figure show $\vphisq^{1/2}$ and
$\vrsq^{1/2} = \vthsq^{1/2}$, respectively. We plot separately each
angular sector of the polar grid in the meridional plane. In each row,
the first panel is closest to the symmetry axis and the last one is
closest to the equatorial plane. Full lines show predictions of the
extended Schwarzschild technique, and dashed lines the results
obtained from the Jeans equations. The model predictions were
interpolated between the $(E,L_z)$ grid points to get smoother
results. Overall the agreement is very good, and better than 1\%.
This is better than the $\sim 5$\% agreement in the DF, because the
velocity moments are integrals over the DF (such that errors tend to
cancel). The errors in the velocity moments are largest near the
symmetry axis, since in the extended Schwarzschild technique only a
few components with very low $L_z$ can reach this region of the
meridional plane. However, the errors are always $\lta 2
\kms$.

\subsection{Reproducing the Test Model with Regular orbits}
\label{ss:testregorb}

The next step in our testing procedure is to try to reproduce the
properties of the test model with regular orbits, rather than
two-integral components. The first obvious question is whether we can
give the orbit superposition algorithm constraints that force it to
generate a model with a DF of the form $f(E,L_z)$, which can then be
compared to the distribution function ${\rm DF}_{\rm HQ}$ calculated
as in Section~\ref{ss:testmodel}. Unfortunately, there is no set of
simple linear kinematic constraints that force the NNLS algorithm to
produce an $f(E,L_z)$ model. One can certainly impose the necessary
conditions that $\vrsq =
\vthsq$ and $\langle v_r v_{\theta} \rangle = 0$, but these conditions
are not sufficient; an $f(E,L_z)$ model is fully determined only by
constraints on {\it all} its higher order velocity moments (e.g.,
Magorrian \& Binney 1994).

We therefore restrict ourselves here to a simpler test. We calculate
an orbit library in the gravitational potential of the test model, but
do not do a subsequent NNLS fit. Instead, we fix the orbital weights
$\gamma_j$ to those appropriate for an $f(E,L_z)$ model, and merely
calculate the projected kinematical quantities for some observational
test setup, given these orbital weights. The results are compared to
the same quantities but now calculated from ${\rm DF}_{\rm HQ}$ as
described in Section~\ref{ss:testmodel}. This tests all of the
important parts of our method that were not already tested by the
calculations in Section~\ref{ss:testtwoint}, namely the orbit
calculation, the projection into data cubes and VPs, and the seeing
convolution.

The main difficulty with this test is that it requires knowledge of
the orbital weights for an $f(E,L_z)$ model, i.e., of the differential
mass density ${\d}M \, / \, {\d}E \, {\d}L_z \, {\d}\Rzvc$, on the
grid of quantities $(E,L_z,\Rzvc)$ that we use to sample orbit space
(cf.~Section~\ref{ss:orbitschoice}). This is not as straightforward as
the calculation of ${\d}M \, / \, {\d}E \, {\d}L_z$ described in
Appendix~\ref{ss:weightstwoint}. In fact, the orbital weights can only
be easily calculated if an explicit expression exists for the third
integral, which is not the case for our test model. However, if the
model has a central BH then at small radii, or high energies, the
potential is Keplerian and spherical (as it is at large radii, or low
energies, because of the finite mass of the model). In this potential,
all the integrals of motion are known, and these limits are therefore
analytically tractable.

At high energies (small radii) the test model reduces to a scale-free
axisymmetric mass density cusp with an $f(E,L_z)$ DF in a spherical
Kepler potential. This limit was studied analytically by de Bruijne,
van der Marel \& de Zeeuw (1996).  In this limit, the normalized
distribution of mass over $(\eta,\Rzvc / R_c)$ at fixed energy, which
we will denote as $h(\eta,\Rzvc/R_c)$, is a known function that is
independent of energy. An explicit expression for $h(\eta,\Rzvc/R_c)$
is derived in Appendix~\ref{ss:weightsKep}. We use this result to
approximate the differential mass density of our test model,
restricting ourselves to the case with a $3 \times 10^6
\Msun$ BH. First, we calculate the differential mass density of the
model over energy alone: $G(E) \equiv {\d}M \, / \, {\d}E = \int 
[{\d}M \, / \, {\d}E \, {\d}L_z] \, {\d}L_z$; where ${\d}M \, / \, {\d}E \,
{\d}L_z$ is obtained from equation~(\ref{gammaptwoint}). Then, we
assume that the distribution of mass over $(L_z,\Rzvc)$ at fixed
energy is always the same as in the high-energy limit, so that
\begin{equation}
{{\d}M \over {{\d}E \, {\d}\eta \, {\d}(\Rzvc/R_c)}}
 = G(E) \> h(\eta,\Rzvc/R_c).
\label{difmass}
\end{equation}
This relation is only correct at asymptotically high energies.  We
found it was sufficiently good at energies with $R_c(E) \lta 0.5''$,
which is where the mass density of the model is a pure power law and
where the potential is Keplerian.

For our test we picked an observational test-setup with the same set
of 8 square apertures, roughly aligned on the major axis, as used by
van der Marel \etal (1997b) in their HST observations of M32 (see
their Figure 3). These apertures all lie at projected radii $R \lta
0.5''$, so most of the light seen in these apertures originates from
stars with energies for which our approximation of the differential
mass distribution is adequate. We chose the same PSF as in van der
Marel \etal (1997b), which is a sum of three Gaussians that
approximates the HST PSF. Subsequently, we picked a grid in
$(E,L_z,\Rzvc)$ space with $N_E=20$, $N_\eta=7$, and $N_{I_3}=7$, and
we calculated an orbit library for this grid. Then finally we
calculated orbital weights from equation~(\ref{difmass}), by
integrating at each point of our $(E,L_z,\Rzvc)$ grid the
approximation ${\d}M / \,[{\d}E \, {\d}\eta \, {\d}(\Rzvc/R_c)]$ over
the corresponding grid cell.  Predictions for the projected kinematics
then follow by superposing the VPs for individual orbits in the
library as in equation~(\ref{basic_Sch_equation}).

Figure~\ref{f:hstsetup} shows the results thus obtained for the
kinematical quantities $V, \sigma, h_3, \ldots, h_6$. As mentioned in
Section~\ref{s:intro}, independent software implementations of the
extended Schwarzschild method were programmed by both N.C.~and
R.v.d.M. Dotted curves in the figure show the results from N.C.'s
software, while dashed curves show the results from R.v.d.M.'s
software. For comparison, solid curves show the results obtained by
direct integration over the known ${\rm DF}_{\rm HQ}$, using the
(completely independent) software of Q95 as described in
Section~\ref{ss:testmodel}. The RMS difference between the different
predictions is $\sim 2 \kms$ in $V$ and $\sigma$, and $\sim 0.01$ in
the Gauss-Hermite moments. Kinematical data typically have larger
observational errors than this, so the numerical accuracy of our
method is entirely adequate for modeling real galaxies.

Finally, let us say a few words about the orbits in the library for
this test model. Figure~\ref{f:orbits_bh} shows the orbits in the
library at the energy corresponding to $R_c =
0.25''$. Figure~\ref{f:orbits_no_bh} shows the orbits at the same
$R_c$, in the same model, but now without a BH. The orbits are all
regular and have a stable periodic parent. The parents were determined
using surfaces-of-section (see also: Richstone 1982; Lees \&
Schwarzschild 1992; Evans 1994; Evans, H\"afner \& de Zeeuw 1997) and
are indicated in the figures.  Most orbits in
Figures~\ref{f:orbits_bh}~and~\ref{f:orbits_no_bh} are tubes, and are
parented by the thin tube. Orbits that are not tubes are indicated. A
minority of the low $\vert \eta \vert$ orbits in
Figure~\ref{f:orbits_no_bh} is parented by higher-order resonances,
such as the 3:2 and 4:3 (being the ratio of the $R$- and
$z$-frequencies of the parent). By contrast, most of the low $\vert
\eta \vert$ orbits in Figure~\ref{f:orbits_bh} is parented by the 1:1 
resonance.  Thus the orbital structure of the models with and without
BH is very clearly different. An analysis of the orbital structure of
these models as a function of the BH mass is beyond the scope of the
present paper, but does seem worth further study.

\section{Concluding Remarks}
\label{s:concl}

In this paper we have described an extension of Schwarzschild's method
for building anisotropic axisymmetric dynamical models of galaxies. We
compute a set of orbits in a given mass model and find the
non-negative superposition of these orbits that best reproduces a set
of (photometric and kinematic) constraints. Our method includes the
full VP shape as kinematic constraint. We parametrize the VP using a
GH expansion so that it is specified by a few numbers. The modeling
method is valid for any kind of parametric (or non-parametric) VP
representation and properly takes into account the observational setup
(seeing convolution, pixel binning, error on each constraint). We
obtain smooth models by imposing a regularization scheme in
integral-space.

R97 have described in detail several aspects of this method and
applied it to the spherical case. However, it is not restricted to
this simple geometry, and we have described here the axisymmetric
extension. The mass model used to compute the orbit library may be
complex: it can have a central density cusp, a stellar disk, a central
black hole or an extended dark halo. Applications of our code to the
flattened systems M32 and NGC 4342 were presented in vdM98 and Cretton
\& van den Bosch (1999).
 
We have also devised a new semi-analytic method for constructing
simpler dynamical models, for which the DF has the special form
$\DF=f(E,L_z)$ or $\DF=f(E)$. These DFs are obtained by using NNLS
with analytic building blocks for which the VPs are obtained by
one-dimensional quadratures. This technique is general and does not
require the density to be expressed analytically as a function of the
potential, but can be used with any complex mass model. Previous
techniques assuming $\DF=f(E,L_z)$ that are also free of this
condition include those of Hunter \& Qian (1993), Dehnen \& Gerhard
(1994), Kuijken (1995), Magorrian (1995) and M96b.

We have tested our new method by having it reproduce the properties of
$f(E,L_z)$ models for which the DF and projected properties can be
calculated with independent algorithms. This allowed us to test all 
aspects of the superposition method, including the sampling of
integral space, orbit calculation, discreteness effects of the spatial
grids, projection into the data cubes, seeing convolution, and the
NNLS algorithm. Tests with only two-intergal components reproduced the
DF with a mass-weighted RMS accuracy of $\lta 5$\%, and the meridional
plane velocity moments to better than $2 \kms$. Tests with a regular
orbit library indicated accuracies in the projected quantities of
$\sim 2 \kms$ in $V$ and $\sigma$, and 0.01 in the GH-moments. All the
tests that we have done indicate that the accuracy of our method is 
adequate for the interpretation of kinematical data obtained
with realistic setups.

Our technique can be extended to triaxiality. Several parts of the
method will be unaltered for this geometry: the use of projected
quantities, e.g., VP$(\xp,\yp,\vlos)$, the fitting procedure, the
seeing convolution, etc. However, the orbital structure is much richer
than in the axisymmetric case. New orbit families appear (e.g., box
orbits) as well as numerous chaotic regions associated with
resonances. During the numerical integration of a trajectory in such a
mass model, all six phase-space coordinates need to be computed,
since there is no azimuthal symmetry. Consequently, the computing
overhead is significantly higher for triaxial geometries.  Work along
these lines is in progress.


\acknowledgments

NC expresses his thanks to Richard Arnold for his help. It is a
pleasure to thank Frank van den Bosch, Eric Emsellem, Hong--Sheng
Zhao, Yannick Copin, Walter Jaffe and Luis Aguilar for fruitful
discussions on the modeling technique. NC acknowledges the hospitality
of MPA (Garching) and Steward Observatory, and thanks D.\ Pfenniger
for discussions at initial stages of this project and visits to Geneva
Observatory. He was financially supported by a grant from the Swiss
government (Etat du Valais) and by an exchange grant from NUFFIC. He
and PTdZ acknowledge travel support from the Leids Kerkhoven Bosscha
Fonds. RPvdM was supported by NASA through grant number \#
GO-05847.01-94A, Hubble Fellowship \# HF-1065.01-94A and an STScI
Institute Fellowship, all awarded by the Space Telescope Science
Institute which is operated by the Association of Universities for
Research in Astronomy, Incorporated, under NASA contract
NAS5-26555. PTdZ acknowledges the hospitality of the Institute for
Advanced Study where this work was initiated.


\clearpage

\appendix

\section{Construction of $f(E,L_z)$- and $f(E)$-components}
\label{s:compapp}

\subsection{Normalization}
\label{ss:compnorm}

To determine the normalization coefficients $C_{[E_0,L_{z,0}]}$ and
$C_{[E_0]}$ in the definitions of the two-integral and isotropic
components (eqs.~[\ref{fdeltatwoint}, \ref{fdeltaE}]), we seek
expressions for the total mass of a single component. The phase-space
volume in cylindrical coordinates is ${\d}^3 {\vec x} \> {\d}^3
{\vec v} = R^2 \> {\d}R \> {\d}\phi \> {\d}z \> {\d}\Rdot \>
{\d}\phidot \> {\d}\zdot$. We use
\begin{equation}
  {\Rdot}^2 = 2 [E - \Phi(R,z)] - {L_z^2 \over R^2} - \zdot^2 , \qquad
  \phidot = L_z / R^2 ,
\label{Rphidot}
\end{equation}
to switch, at fixed $(R,z)$, from the variables $(\Rdot,\zdot)$ to
$(E,L_z)$. We then find for the total mass of a two-integral
component:
\begin{equation}
  m_{[E_0,L_{z,0}]} \equiv 
        \int {\d}^3 {\vec x} \> {\d}^3 {\vec v} \>
             f^{\delta}_{[E_0,L_{z,0}]}  =
        \int R^2 {\d}R  \int {\d}\phi  \int {\d}z 
        \int {{{\d}E} \over {\Rdot}}  \int {{{\d}L_z} \over {R^2}}  
        \int {\d}\zdot \> f^{\delta}_{[E_0,L_{z,0}]}  .
\label{masstwointdef}
\end{equation}
Upon substitution of $f^{\delta}_{[E_0,L_{z,0}]}$ from
equation~(\ref{fdeltatwoint}), the integration over $\phi$, $E$ and
$L_z$ becomes trivial, and we obtain
\begin{equation}
  m_{[E_0,L_{z,0}]} =
      4 \pi C_{[E_0,L_{z,0}]} 
            \dint {\d}R \> {\d}z
            \int \Bigl \lbrace
                   2 \, [E_0 - \Phi(R,z)] - {L_{z,0}^2 \over R^2} - \zdot^2
                 \Bigr \rbrace^{-1/2} \> {\d}\zdot .
\label{masstwointinterm}
\end{equation}
The integral over ${\d}\zdot$ extends over the region $|\zdot| \leq
\zdot_{\rm max}$, where $\zdot_{\rm max}$ is defined as the root
of the expression in the square root. Therefore,
\begin{equation}
  m_{[E_0,L_{z,0}]} =
      8 \pi C_{[E_0,L_{z,0}]}
            \dint {\d}R \> {\d}z
            \int_{0}^{\zdot_{\rm max}} 
              { {{\d}\zdot} \over 
                {\sqrt{\zdot_{\rm max}^2 - \zdot^2}} }  =
      4 \pi^2 C_{[E_0,L_{z,0}]} \dint {\d}R \> {\d}z ,
\label{masstwoint}
\end{equation}
where the remaining double integral is over the region for which $E_0
- \Phi(R,z) - (L_{z,0}^2/2R^2) \geq 0$. This is exactly the region
$\Phi_{{\rm eff},0}(R,z) \leq E_0$ bounded by the ZVC at the given
$(E_0,L_{z,0})$, where $\Phi_{{\rm eff},0}$ is the effective
gravitational potential at the given $L_{z,0}$. To obtain
$m_{[E_0,L_{z,0}]} = 1$, we choose
\begin{equation}
  C_{[E_0,L_{z,0}]} =
    \Bigl [ 4 \pi^2 \!\!\!\!\!\!\!\!
               \int\limits_{\>\>\>\Phi_{{\rm eff},0}(R,z) \leq E_0}
               \!\!\!\!\!\!\!\!\!\!\!\!\!\!\!\!\int {\d}R \> {\d}z
    \Bigr ]^{-1}    =
    \Bigl [ 2 \pi^2 \!\!\!\!\!
                    \oint\limits_{{\rm ZVC}[E_0,L_{z,0}]}
                    \!\!\!\!\!
         ( R\>{\d}z - z\>{\d}R )
    \Bigr ]^{-1}    ,
\label{Atwoint}
\end{equation}
where the second equality was obtained with a variant of Stokes' theorem. 

Following similar arguments, we obtain for the mass of an isotropic
component:
\begin{eqnarray}
  m_{[E_0]} & \equiv &
        \int {\d}^3 {\vec x} \> {\d}^3 {\vec v} \>
             f^{\delta}_{[E_0]}   \\ \nonumber
            & = &
        8 \pi C_{[E_0]} 
          \dint {\d}R \> {\d}z  \int {\d}L_z  
          \int_{0}^{\zdot_{\rm max}} 
            { {{\d}\zdot} \over 
              {\sqrt{\zdot_{\rm max}^2 - \zdot^2}} }  =
        4 \pi^2 C_{[E_0]} 
               \!\!\!\!\!\!\!\!\!\!\!\!
               \int\limits_{\>\>\>\>\>\Phi_{\rm eff}(R,z) \leq E_0}
               \!\!\!\!\!\!\!\!\!\!\!\!\!\!\!\!\!\dint 
               {\d}R \> {\d}z \> {\d}L_z ,
\label{massisodef}
\end{eqnarray}
where the effective gravitational potential $\Phi_{\rm eff}(R,z)$ is
a function of $L_z$, at fixed $(R,z)$. Evaluation of the integral over 
${\d}L_z$ yields
\begin{equation}
  m_{[E_0]} = 8 \sqrt{2} \> \pi^2 C_{[E_0]} \!\!\!
                  \int\limits_{\>\Phi(R,z) \leq E_0}
                  \!\!\!\!\!\!\!\!\!\!\!\!\!\!\!\!\int 
                  R \sqrt{E_0-\Phi(R,z)} \>\> {\d}R \> {\d}z .
\label{massiso}
\end{equation}
To obtain $m_{[E_0]} = 1$, we choose
\begin{equation}
  C_{[E_0]} =
    \Bigl [ 8 \sqrt{2} \> \pi^2 \!\!\!
                  \int\limits_{\>\Phi(R,z) \leq E_0}
                  \!\!\!\!\!\!\!\!\!\!\!\!\!\!\!\!\int
                  R \sqrt{E_0-\Phi(R,z)} \>\> {\d}R \> {\d}z 
    \Bigr ]^{-1}    .
\label{Aiso}
\end{equation}
For the special case of a spherical potential, $\Phi = \Phi(r)$, we 
have
\begin{equation}
  m_{[E_0]} = 16 \sqrt{2} \> \pi^2 C_{[E_0]} \!\!
                  \int\limits_{\>\Phi(r) \leq E_0}  \!\!\!\!\!\
                  r^2 \sqrt{E_0-\Phi(r)} \>\> {\d}r ,
\label{massisospher}
\end{equation}
which can be recognized as the `density-of-states' function for an
isotropic spherical system (BT). Calculations for
spherical components with DFs proportional to $\delta(E-E_0) \>
\delta(L-L_0)$ were presented in Appendix~A of R97.

\subsection{Velocity profiles}
\label{ss:compjacob}

We derive here the Jacobian $J$ for the transformation from
$(v_{\xp},v_{\yp})$ to $(E,L_z)$, which enters into the expressions
for the VPs of two-integral and isotropic components
(eqs.~[\ref{VPtwoint},\ref{VPiso}]). The energy is
\begin{equation}
  E = \onetwo (v_{\xp}^2 + v_{\yp}^2 + \vlos^2) +
      \Phi (\xp,\yp,\zp) . 
\label{Edef}
\end{equation}
For inclination angle $i$,
\begin{equation}
  x = -\yp \cos i + \zp \sin i   , \qquad
  z =  \yp \sin i + \zp \cos i   ,
\label{xzdef}
\end{equation}
and the angular momentum is therefore
\begin{equation}
  L_z = x v_y - y v_x  
      = v_{\xp} (-\yp \cos i + \zp \sin i) + 
        v_{\yp} (\xp \cos i) - 
        \vlos (\xp \sin i) .
\label{Lzdef}
\end{equation}
This yields for the Jacobian
\begin{equation}
  J = \Bigl | \xp v_{\xp} \cos i + 
              \yp v_{\yp} \cos i -
              \zp v_{\yp} \sin i 
      \Bigr |^{-1} ,
\label{generaljacobian}
\end{equation}
in which $v_{\xp}$ and $v_{\yp}$ are functions of $E$ and $L_z$
determined by:
\begin{equation}
  v_{\xp} = { {L_z - v_{\yp} \xp \cos i + \vlos \xp \sin i} \over 
              {-\yp \cos i + \zp \sin i} }    , \qquad
  v_{\yp}^2 = 2(E-\Phi) - v_{\xp}^2 - \vlos^2 .
\label{vxyproj}
\end{equation}
Substitution of $v_{\xp}$ in the expression for $v_{\yp}^2$ yields 
a quadratic equation for $v_{\yp}$:
\begin{equation}
  a v_{\yp}^2 + b v_{\yp} + c = 0, 
\label{vypabc}
\end{equation}
where 
\begin{eqnarray}
  a &=& (-\yp \cos i + \zp \sin i)^2 + (\xp \cos i)^2  , \nonumber \\
  b &=& -2 (L_z + \vlos \xp \sin i) \> \xp \cos i      , \nonumber \\
  c &=& \vlos^2 (-\yp \cos i + \zp \sin i)^2 + 
          (L_z + \vlos \xp \sin i)^2 - 
          2(E-\Phi)(-\yp \cos i + \zp \sin i)^2        . 
\label{abcdef}
\end{eqnarray}
Therefore,
\begin{equation}
  v_{\yp} = { {2 (L_z + \vlos \xp \sin i) \>
                 \xp \cos i  \pm \sqrt{\Delta} } \over
              {2 [ (-\yp \cos i + \zp \sin i)^2 + 
                   (\xp \cos i)^2] } }  ,
\label{vypsolution}
\end{equation}
with $\Delta \equiv b^2 - 4ac$. Equations~(\ref{generaljacobian}),
(\ref{vxyproj}) and~(\ref{vypsolution}) define the Jacobian $J$.

\section{Relation Between Orbital Weights and the DF}
\label{s:weightsDF}

\subsection{${\d}M \, / \, {\d}E \, {\d}L_z$ for an $f(E,L_z)$ model}
\label{ss:weightstwoint}

For a two-integral model there is a simple relation between the
differential mass density ${\d}M \, / \, {\d}E
\, {\d}L_z$ and the DF $f(E,L_z)$. To derive this relation 
(see also Vandervoort 1984) we write the trivial identity
\begin{equation}
  f(E,L_z) = \dint f(E_0,L_{z,0}) \> 
               ( f^{\delta}_{[E_0,L_{z,0}]} / C_{[E_0,L_{z,0}]} ) \>
               {\d}E_0 \> {\d}L_{z,0} ,
\label{DFtwointdeltaint}
\end{equation}
where $f^{\delta}_{[E_0,L_{z,0}]}$ and $C_{[E_0,L_{z,0}]}$ are as
defined in equation~(\ref{fdeltatwoint}). The total mass of the system is
\begin{equation}
  M \equiv \int {\d}^3 {\vec x} \> {\d}^3 {\vec v} \> f(E,L_z)
    = \dint f(E_0,L_{z,0}) \> ( m_{[E_0,L_{z,0}]} / C_{[E_0,L_{z,0}]} ) \>
            {\d}E_0 \> {\d}L_{z,0} ,
\label{Mtwoint}
\end{equation}
where the second identity follows upon substitution of
equation~(\ref{DFtwointdeltaint}), exchange of the integration order,
and use of the definition of $m_{[E_0,L_{z,0}]}$ from
equation~(\ref{masstwointdef}). Substitution of
equations~(\ref{masstwoint}) and~(\ref{Atwoint}), relabeling of the
integration variables from $(E_0,L_{z,0})$ to $(E,L_z)$, and
differentiation then yields
\begin{equation}
  {\d}M \> / \> {\d}E \> {\d}L_z = 
      f(E,L_z) \> \times \> 
      \Bigl [
         2 \pi^2 \!\!\!\!\!
                 \oint\limits_{{\rm ZVC}[E,L_z]} 
                 \!\!\!\!\!
         ( R\>{\d}z - z\>{\d}R )
      \Bigr ]  .
\label{gammaptwoint}
\end{equation}
The mass weight $\gamma_j$ for a cell in integral space is
\begin{equation}
  \gamma_j = 
  \int\limits_{\null\>{\rm cell}_j} \!\!\!\!\!\int
  {{\d}M \over {\d}E \, {\d}L_z} \; {\d}E\,{\d}L_z.
\label{gamma_j_mass}
\end{equation}

\subsection{Scale-free Density in a Kepler Potential}
\label{ss:weightsKep}

We summarize here the asymptotic case of a scale-free axisymmetric
density in a spherical Kepler potential, which was discussed in detail
by de Bruijne, van der Marel \& de Zeeuw (1996). We adopt the same
units as in that paper. In those units, $\rho_{\star} = s^{-\xi}$ and
$\Psi = 1 / r$, where $s$ is defined as in eq.~(\ref{densitylaw}) by
$s^2 = R^2 + (z/q)^2$, with $q$ the axial ratio. The associated
eccentricity is $e = \sqrt{1-q^2}$. It is convenient to work with the
integrals of motion
\begin{equation}
  {\cal E} = - E                                  , \qquad
  {\eta}^2 \equiv L_z^2 / L_{\rm max}^2({\cal E}) , \qquad
  {\zeta}^2 \equiv L^2 / L_{\rm max}^2({\cal E})  , 
\label{zetaetadef}
\end{equation}
where ${\cal E}$ is the binding energy, $\eta^2 \in [0,1]$ and
$\zeta^2 \in [\eta^2,1]$.  The unique (even) two-integral DF is
\begin{equation}
  \DF ({\cal E},\eta^2) = C_0 \> g({\cal E}) \> j(e^2 \eta^2) ,
\label{fasymp}
\end{equation}
where
\begin{equation}
  C_0 = { {q^{\xi}} \over
          {2 \pi \> {\cal B}(\xi\!-\!{1\over2},{3\over2})} } , \qquad
  g({\cal E}) = {\cal E}^{\xi-{3\over 2}}                    , \qquad
  j (e^2 \eta^2) =
      {}_3F_2 ( \fr{\xi}{2}, \fr{\xi+1}{2}, \fr{\xi+2}{2};
                    \fr{1}{2}, \fr{2\xi - 1}{2}; e^2 \eta^2) .
\label{fasympfac}
\end{equation}
The special functions ${\cal B}$ and ${}_3F_2$ are the beta-function
and a generalized hypergeometric function, respectively.  Upon
substitution of $\xi = -\alpha$ this yields the $R_c \rightarrow 0$
limit of our test model (Section~\ref{ss:testmodel}); upon
substitution of $\xi = -\alpha - (\beta\gamma) - (\delta\epsilon)$, it
yields the $R_c \rightarrow \infty$ limit.

We wish to calculate the mass weight $\gamma_j$ 
contained in a cell number $j$ of integral space (see 
eq.~[\ref{basic_Sch_equation}]). According to
equations~(35) and ~(37) of de Bruijne et al., this is given by:
\begin{equation}
  \gamma_j = 
     \int \!\!\!\!\int\limits_{{\rm cell}_j} 
          \!\!\!\!\int
     {d}{\cal E} \> {\d}\eta^2 \> {\d}\zeta^2 \> 
     w({\cal E},\eta^2,\zeta^2) \> \DF({\cal E},\eta^2) , \qquad
     w ({\cal E},\eta^2,\zeta^2) \> = \>
     {{\pi^3}\over{4}} \> {\cal E}^{-5/2} \> (\eta^2)^{-1/2} \>
     (\zeta^2)^{-1/2}  .
\label{deltaMdef}
\end{equation}
In the Kepler potential the binding energy is related to the circular
radius according to ${\cal E} = 1 / (2 R_c)$. The ZVC at a given
$(R_c,\eta^2)$ is therefore defined by
\begin{equation}
  1 = { {2R_c}\over {r}} - ({{R_c \eta}\over {R}})^2   .
\label{ZVCasymp}
\end{equation}
A particle on the ZVC has $v_r = v_{\theta} = 0$, $L^2 = r^2
v_{\phi}^2$ and $L_z^2 = R^2 v_{\phi}^2$. Therefore, we have $\zeta^2 = (r
\eta / R)^2$. Combined with the expression for the ZVC this yields
\begin{equation}
  \zeta^2 = \Bigl [ { {2 \eta \> (\Rzvc/R_c)} \over
                      {\eta^2 + (\Rzvc/R_c)^2}  } 
            \Bigr ]^2 ,
\label{zetaRzvc}
\end{equation}
which is a one-to-one relation if $(\Rzvc/R_c)$ is chosen between
$\eta$ and $1 + \sqrt{1-\eta^2}$. Substitution of the DF from
equation~(\ref{fasymp}) into equation~(\ref{deltaMdef}), and
transformation to the variables $\log R_c$, $\eta \in [0,1]$ and
$(\Rzvc/R_c) \in [\eta, \> 1 + \sqrt{1-\eta^2}]$ yields
\begin{eqnarray}
    { {{\d}M} \over { {\d}(\log R_c) \, {\d}\eta \, {\d}(\Rzvc/R_c) } } =
        2 \, C_0 \> \pi^3 \> ({{R_c}\over {2}})^{3-\xi} \> 
        \eta \> j(e^2 \eta^2) \> 
        \Bigr \lbrace 
           { {\eta^2 - (\Rzvc/R_c)^2} \over
             {[\eta^2 + (\Rzvc/R_c)^2]^2} } 
        \Bigl \rbrace                             .
\label{deltaMasymp}
\end{eqnarray}
Hence, the normalized distribution of mass over $(\eta,\Rzvc / R_c)$
at fixed energy, which we will denote as $h(\eta,\Rzvc/R_c)$, is
independent of energy. In particular:
\begin{eqnarray}
  h(\eta,\Rzvc/R_c) 
      &=& 2 \eta \> j (e^2 \eta^2) \>
        \Bigr \lbrace
           { {\eta^2 - (\Rzvc/R_c)^2} \over
             {[\eta^2 + (\Rzvc/R_c)^2]^2} }
        \Bigl \rbrace                            \> \Bigl / \nonumber \\
      &&  \>\>\> \int_0^1 \d{\eta^2} \> j (e^2 \eta^2)  
	    \int_{\eta}^{1 + \sqrt{1-\eta^2}} \d (\Rzvc/R_c) \>
        \Bigr \lbrace
           { {\eta^2 - (\Rzvc/R_c)^2} \over
             {[\eta^2 + (\Rzvc/R_c)^2]^2} }
        \Bigl \rbrace                             .
\label{ME2}
\end{eqnarray}

\clearpage


{}

\clearpage


\ifsubmode\else
\baselineskip=14pt
\fi


\newcommand{\figcapflowchart}{Flowchart of the extended 
Schwarzschild method. We find the non-negative superposition of the
orbits with a least squares algorithm (NNLS). This combination of
orbits reproduces a set of photometric (surface brightness
distribution) and kinematic constraints (VPs).
\label{f:flowchart}}

\newcommand{\figcapfigZVCs}{Zero Velocity Curves are plotted for 
7 values of $L_z$ uniformly sampled between $0.05 L_{z}^{\rm max}$ and
$0.95 L_{z}^{\rm max}$ at an energy corresponding to a circular orbit
radius $R_c=2.0$ (indicated by the dot). The mass model is the one of
our test model of Section~\ref{ss:testmodel}, with no BH.
\label{f:fig_ZVCs}}

\newcommand{\figcapanorbit}{A regular orbit, the thin-tube periodic 
orbit, and the ZVC around them in the meridional $(R,z)$ plane for the
same mass model as in Figure~\ref{f:fig_ZVCs}. The radii ${\rm R}_{\rm
min}$, ${\rm R}_{\rm max}$ and ${\rm R}_{\rm thin}$ are indicated (see
text). The circular orbit is represented by a black dot. The different
starting points on the ZVC are shown with open dots and the
corresponding angles $w$ and $w_{\rm thin}$ are indicated.
\label{f:anorbit}}

\newcommand{\figcaporbitalVPs}{VPs as a function of line-of-sight velocity
$\vlos$ (in $\kms$) for the two orbits of Figure~\ref{f:anorbit}. The
regular and thin orbits are shown in the top and bottom panels,
respectively, for viewing through ($1''$-square) cells along the major
axis (left) and minor axis (right), respectively. The orbits were not
convolved with a PSF. For each panel, different lines correspond to
cells at different distances $r$ from the center: the full line
corresponds to the central cell ($r=0''$), the dotted line to $r=1''$,
the short dashed line to $r=2''$ and the dot-short dashed line to
$r=3''$.\label{f:orbitalVPs}}

\newcommand{\figcaptwointVPs}{VPs of an individual two-integral component 
with the same $(E,L_z)$ as the orbits in Figure~\ref{f:anorbit},
viewed along the major axis (left) and minor axis (right),
respectively. Line types are the same as in
Figure~\ref{f:orbitalVPs}. However, unlike in
Figure~\ref{f:orbitalVPs}, these VPs were calculated for a point along
either axis, and were not `integrated' over cells. \label{f:twointVPs}}

\newcommand{\figcapDFreconstructionBH}{The DF $f(E,L_z)$ for 
the test model with a $3\times 10^6 \Msun$ BH described in
Section~\ref{ss:testmodel}. The DF from the semi-analytical HQ
algorithm and from the extended Schwarzschild technique, using only
two-integral components, are plotted as function of the component
number. The components run in order or energy, and in order of $L_z$
for each energy; this causes the jagged appearance of the curves. The
two curves mostly overlap in the comparison interval (dashed
lines). The insert shows the relative difference (in per cent) between
the two DFs. The agreement between the two methods of calculating the
DF is acceptable.\label{f:DF_reconstruction_BH}}

\newcommand{\figcapDFreconstructionnoBH}{Similar to 
Figure~\ref{f:DF_reconstruction_BH}, but for the same model without a
central BH.\label{f:DF_reconstruction_no_BH}}

\newcommand{\figcaprmscombined}{The RMS logarithmic residual 
$\RMS_{\log \DF}$ for the test model of
Figure~\ref{f:DF_reconstruction_no_BH}, as function of the logarithm
of the regularization parameter $\Delta$. The residual measures the
difference between the DF as calculated with the semi-analytical HQ
algorithm, and as calculated with the extended Schwarzschild
technique, using only two-integral components. The different
line-types indicate the cases in which only masses in the meridional
plane are included as consistency constraints (dotted line), in which
only projected plane masses are included (dashed line), or in which
both are included (full line with dots).\label{f:rms_combined}}

\newcommand{\figcapmomentsnoBH}{Comparison of meridional plane 
velocity moments, calculated either with the extended Schwarzschild
technique using only two-integral components (full curves), or with
the Jeans equations (dashed curves), for the model of
Figure~\ref{f:DF_reconstruction_no_BH}. The top and bottom row of
panels show $\vphisq^{1/2}$ and $\vrsq^{1/2} = \vthsq^{1/2}$,
respectively. The meridional $(r,\theta)$ plane is divided in 7
sectors. In each row, the first panel is the sector closest to the
symmetry axis and the last panel is the sector closest to the
equatorial plane. The discrepancies are largest near the symmetry
axis, but are acceptable everywhere.\label{f:2nd_moments_no_BH}}

\newcommand{\figcaphstsetup}{Kinematical predictions for the edge-on
$f(E,L_z)$ test model with a $3 \times 10^6 \Msun$ BH discussed in
Section~\ref{ss:testregorb}. The kinematical apertures are the same as
for the HST observations of M32 by van der Marel et al (1997b). They
are aligned along the major axis. Data points are plotted
equidistantly along the abscissa. Dotted and dashed curves are
predictions obtained with the extended Schwarzschild technique, using
the software of Cretton and van der Marel, respectively. The solid
curves show predictions obtained from direct integration over the DF,
using the software of Q95. The three curves agree well, demonstrating
the numerical accuracy of the orbit superposition
technique.\label{f:hstsetup}}

\newcommand{\figcaporbitsbh}{Examples of orbits at the energy with 
$R_c = 0.25''$, in the test model of Section~\ref{ss:testmodel} for
the case with a $3 \times 10^6 \Msun$ BH. The axes for each orbit are
in units of $R_c$.  Each line corresponds to a different value of
$\vert\eta\vert$ and each column to a different value of the third
integral. The ratio of the $R$- and $z$-frequencies for the parent
orbit is indicated for those orbits that are not parented by the thin
tube.\label{f:orbits_bh}}

\newcommand{\figcaporbitsnobh}{Similar as Figure~\ref{f:orbits_bh}, 
but now for the same model without a central BH.\label{f:orbits_no_bh}}


\ifsubmode
\figcaption{\figcapflowchart}
\figcaption{\figcapfigZVCs}
\figcaption{\figcapanorbit}
\figcaption{\figcaporbitalVPs}
\figcaption{\figcaptwointVPs}
\figcaption{\figcapDFreconstructionBH}
\figcaption{\figcapDFreconstructionnoBH}
\figcaption{\figcaprmscombined}
\figcaption{\figcapmomentsnoBH}
\figcaption{\figcaphstsetup}
\figcaption{\figcaporbitsbh}
\figcaption{\figcaporbitsnobh}

\clearpage
\else\printfigtrue\fi

\ifprintfig


\clearpage
\begin{figure}
\epsscale{1.0}
%
\vskip1.5truecm
\plotone{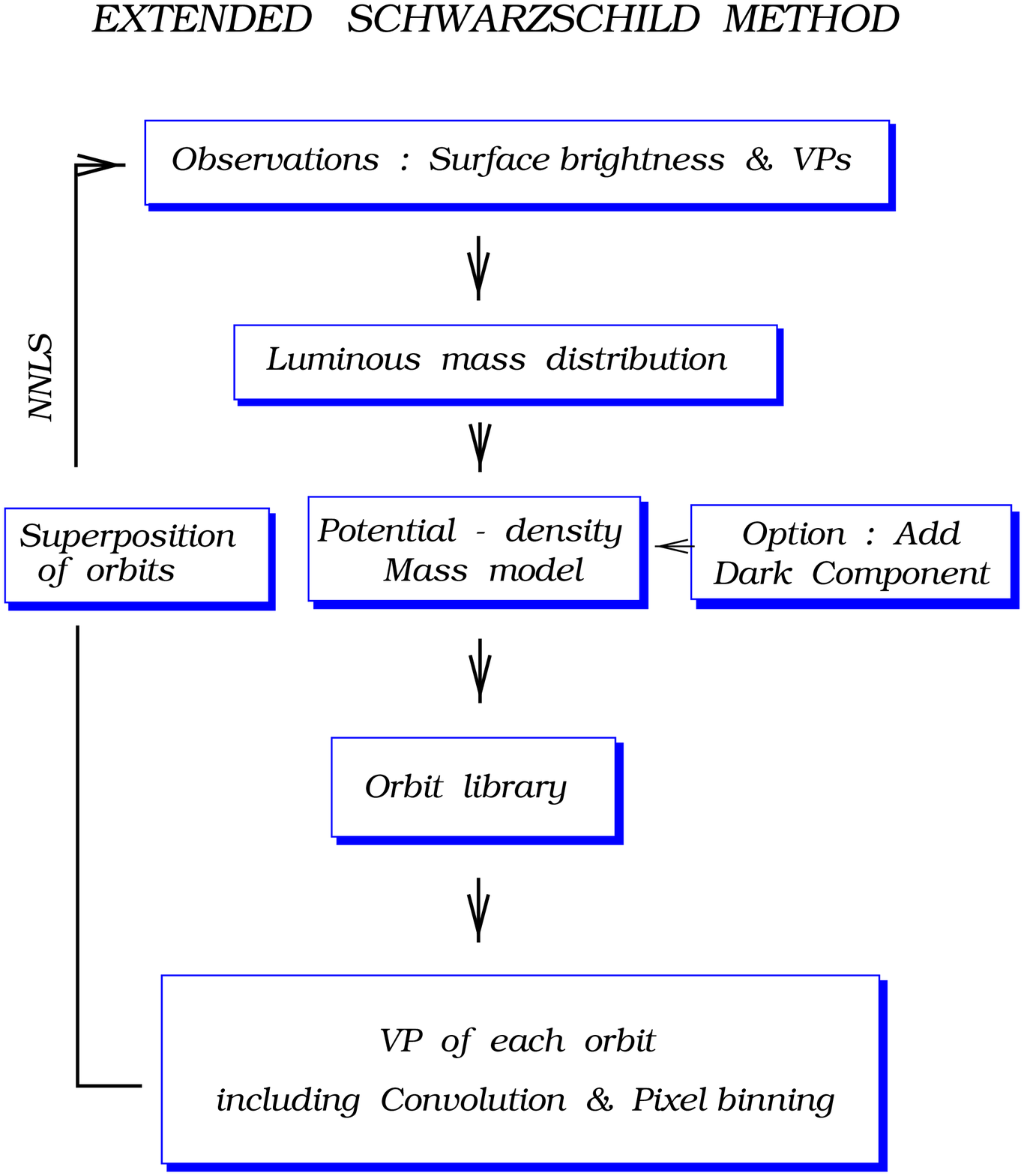}
\vskip-0.5truecm
\ifsubmode
\setcounter{figure}{0}
\addtocounter{figure}{1}
\centerline{Figure~\thefigure}
\else\figcaption{\figcapflowchart}\fi
\end{figure}


\clearpage
\begin{figure} 
\epsscale{1.} 
\plotone{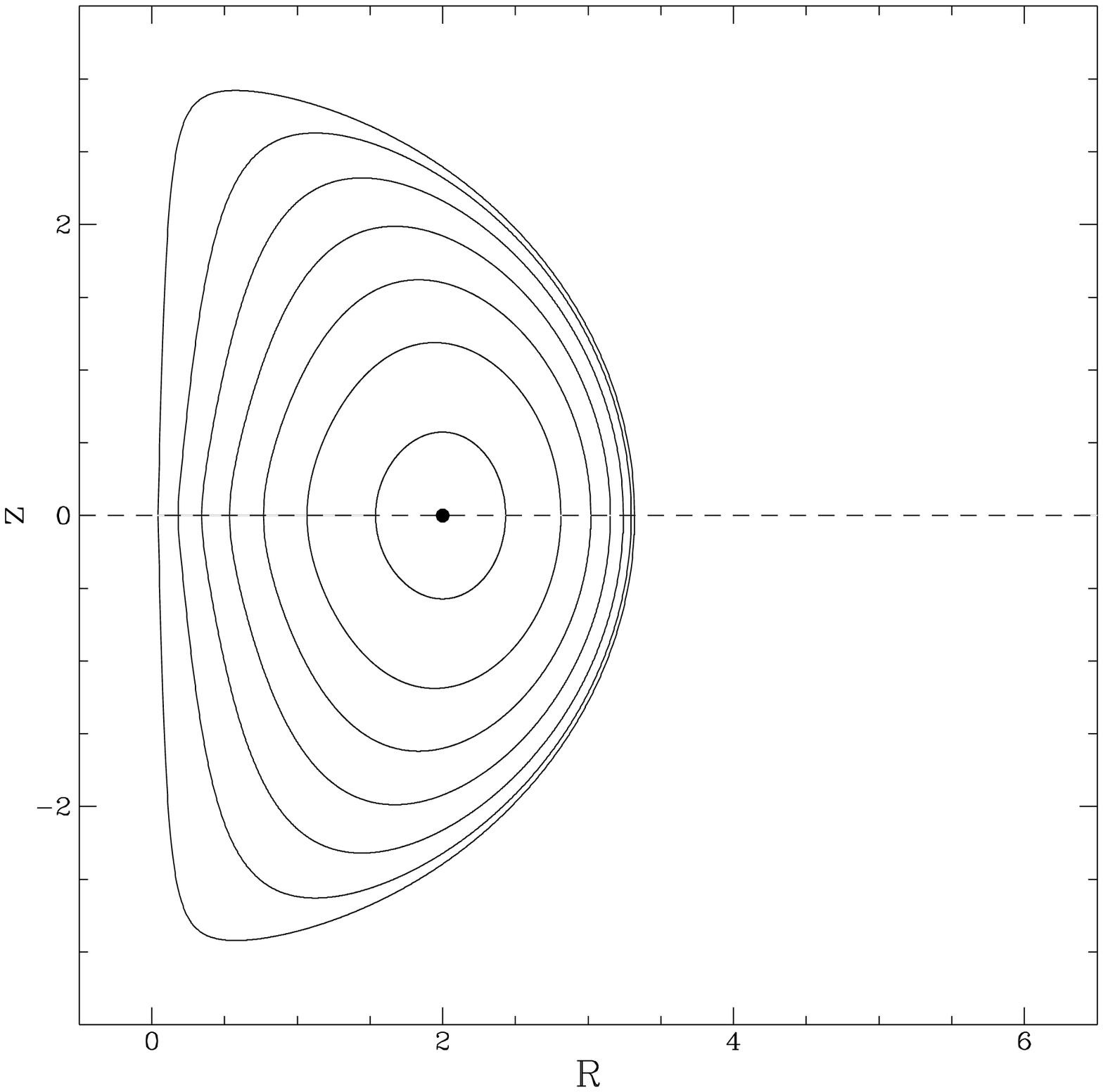}
\ifsubmode
\vskip3.0truecm
\addtocounter{figure}{1}
\centerline{Figure~\thefigure}
\else\figcaption{\figcapfigZVCs}\fi
\end{figure}


\clearpage
\begin{figure} 
\epsscale{1.} 
\plotone{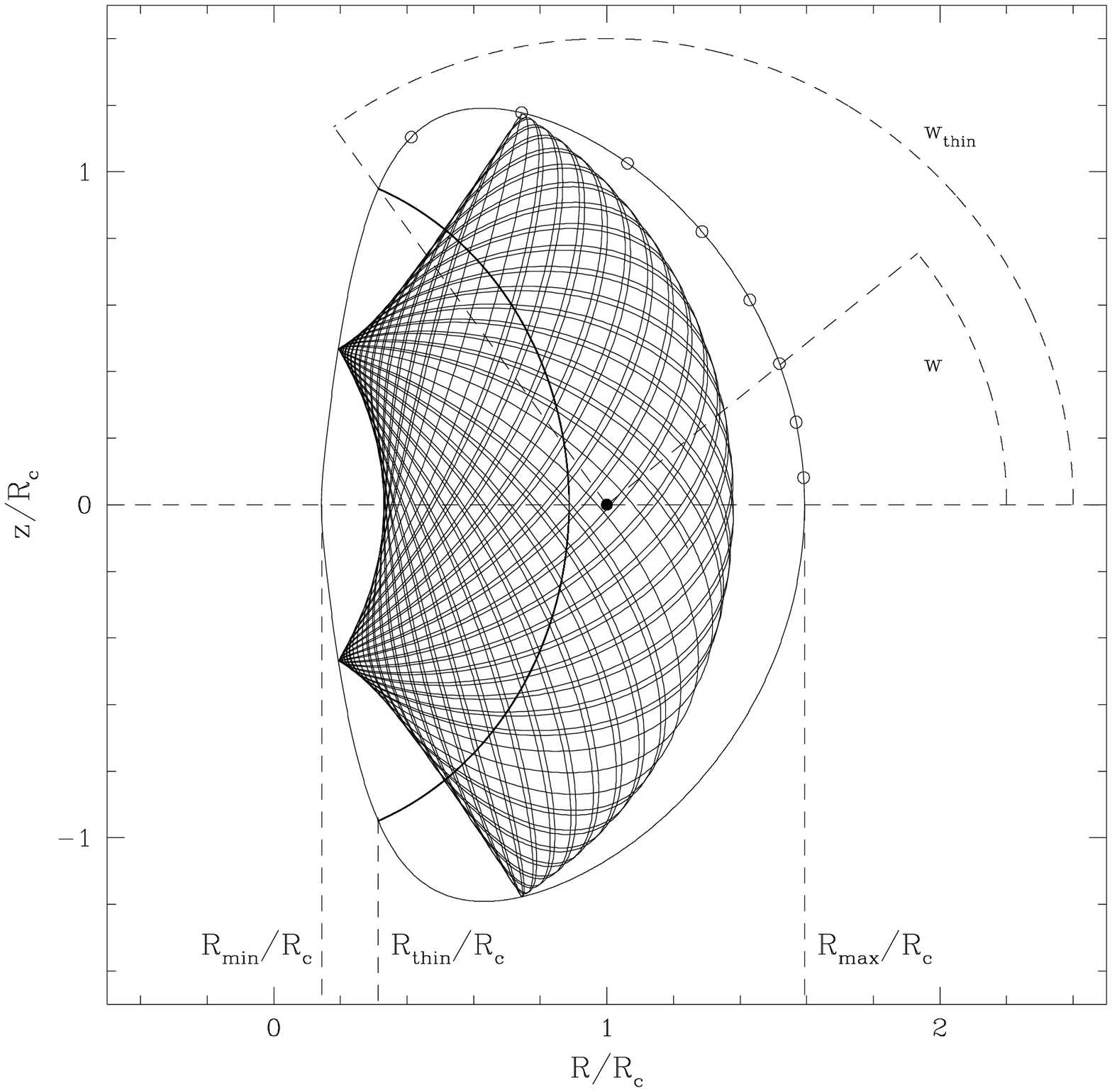}
\ifsubmode
\vskip3.0truecm
\addtocounter{figure}{1}
\centerline{Figure~\thefigure}
\else\figcaption{\figcapanorbit}\fi
\end{figure}


\clearpage
\begin{figure} 
\epsscale{0.75} 
\plotone{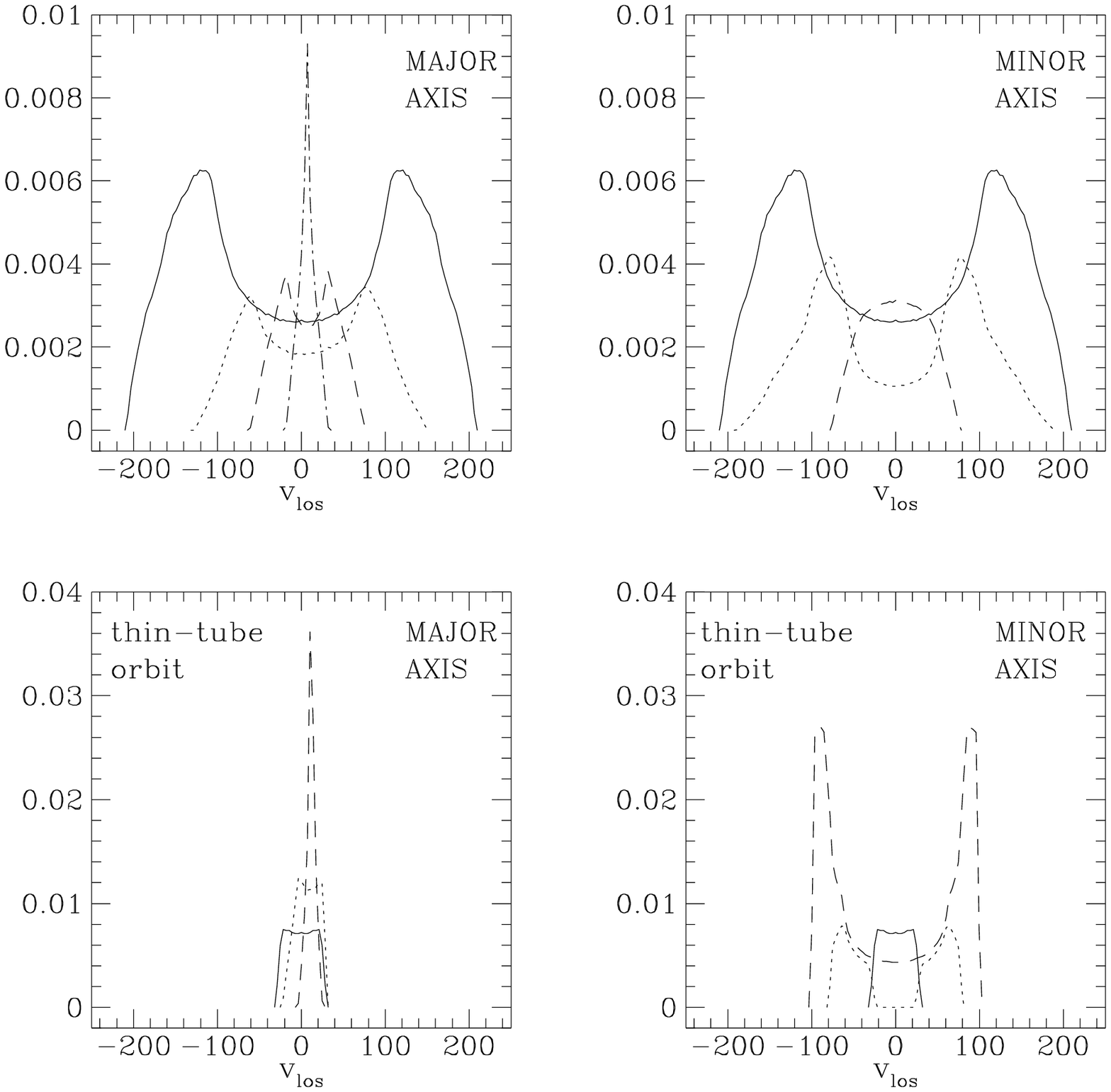}
\ifsubmode
\vskip3.0truecm
\addtocounter{figure}{1}
\centerline{Figure~\thefigure}
\else\figcaption{\figcaporbitalVPs}\fi
\end{figure}


\clearpage
\begin{figure} 
\epsscale{0.75}
\plotone{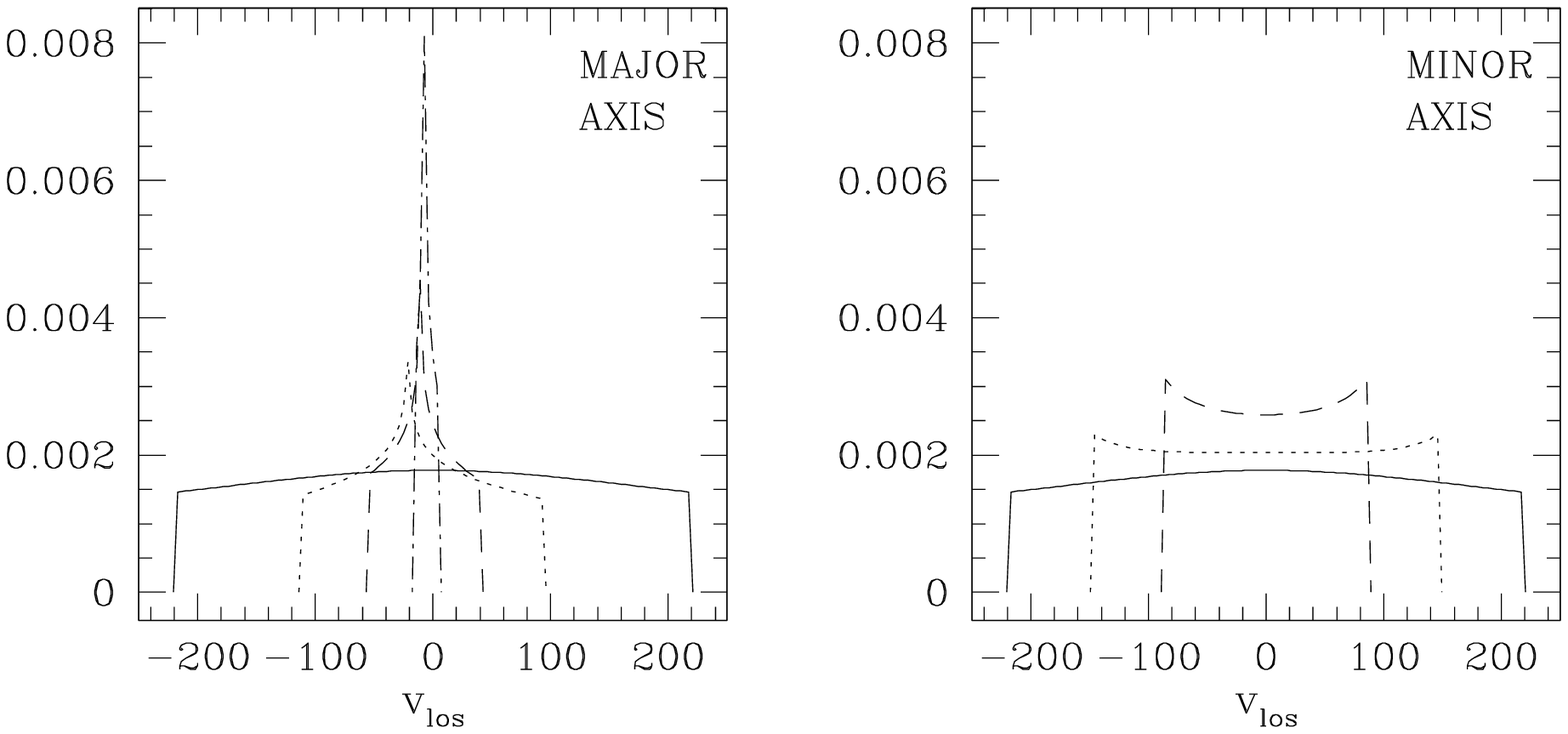}
\ifsubmode
\vskip3.0truecm
\addtocounter{figure}{1}
\centerline{Figure~\thefigure}
\else\figcaption{\figcaptwointVPs}\fi
\end{figure}


\clearpage
\begin{figure} 
\epsscale{1.0}
\plotone{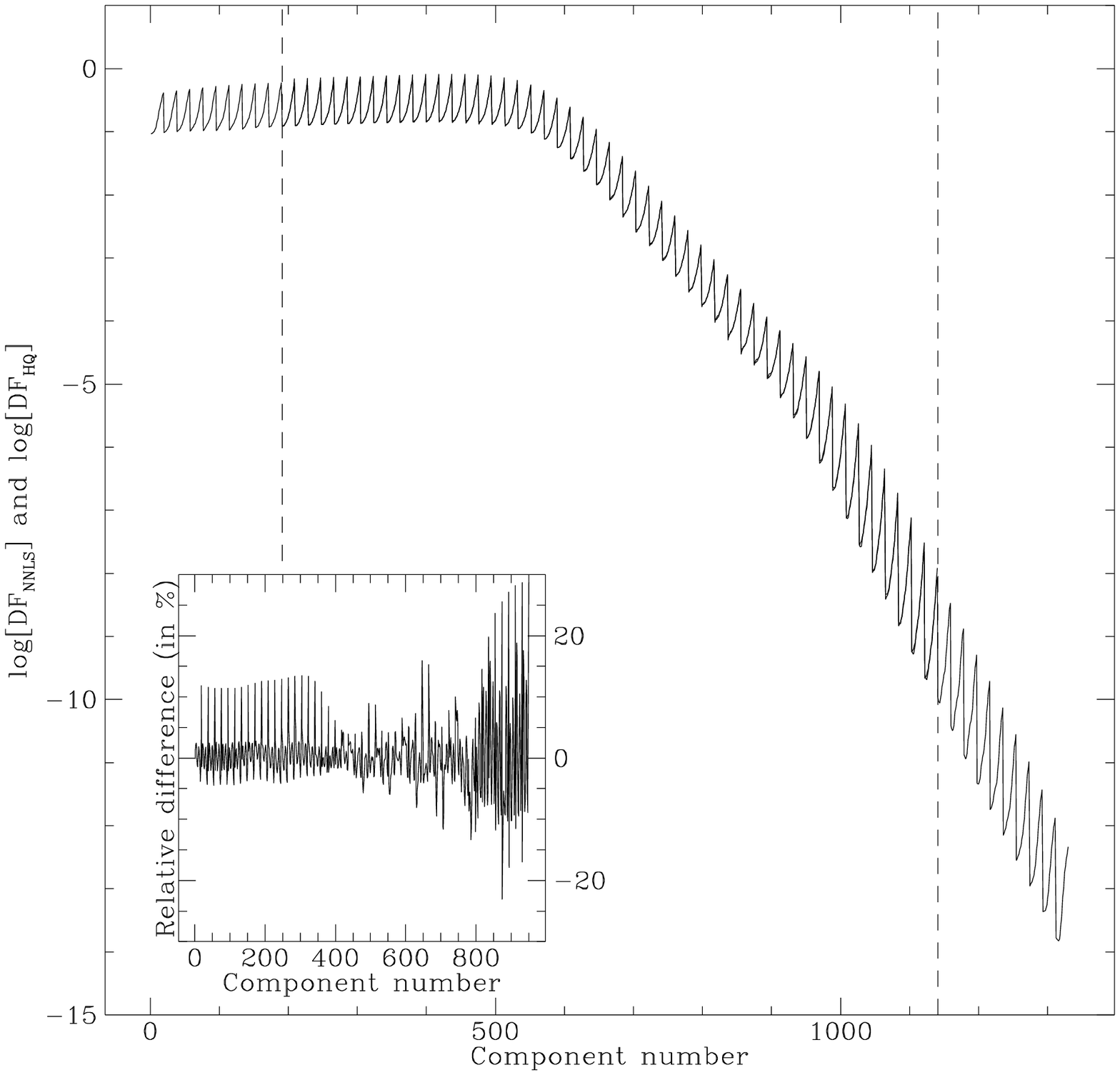}
\ifsubmode
\vskip3.0truecm
\addtocounter{figure}{1}
\centerline{Figure~\thefigure}
\else\figcaption{\figcapDFreconstructionBH}\fi
\end{figure}


\clearpage
\begin{figure} 
\epsscale{1.0}
\plotone{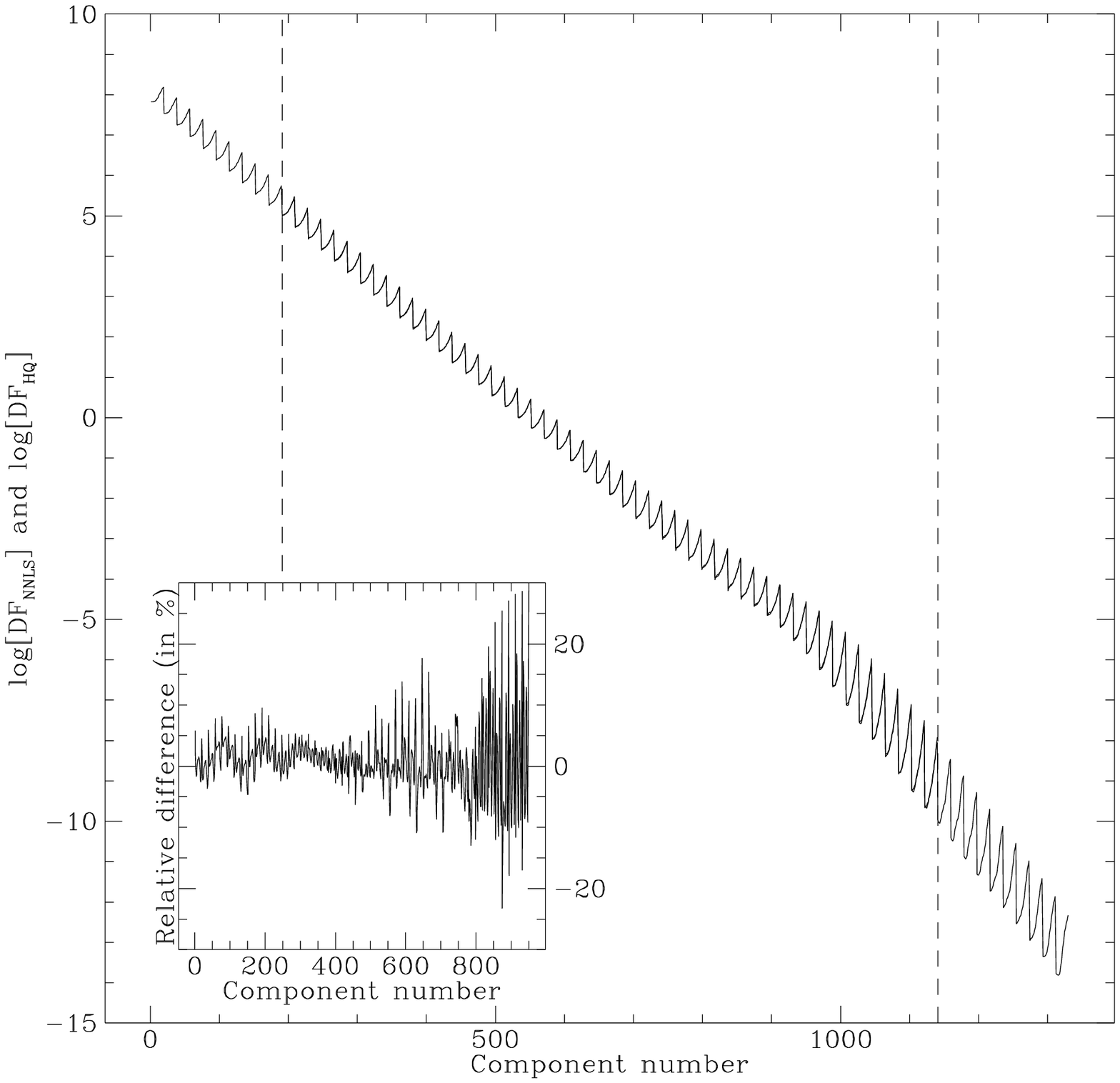}
\ifsubmode
\vskip3.0truecm
\addtocounter{figure}{1}
\centerline{Figure~\thefigure}
\else\figcaption{\figcapDFreconstructionnoBH}\fi
\end{figure}


\clearpage
\begin{figure} 
\epsscale{1.0}
\plotone{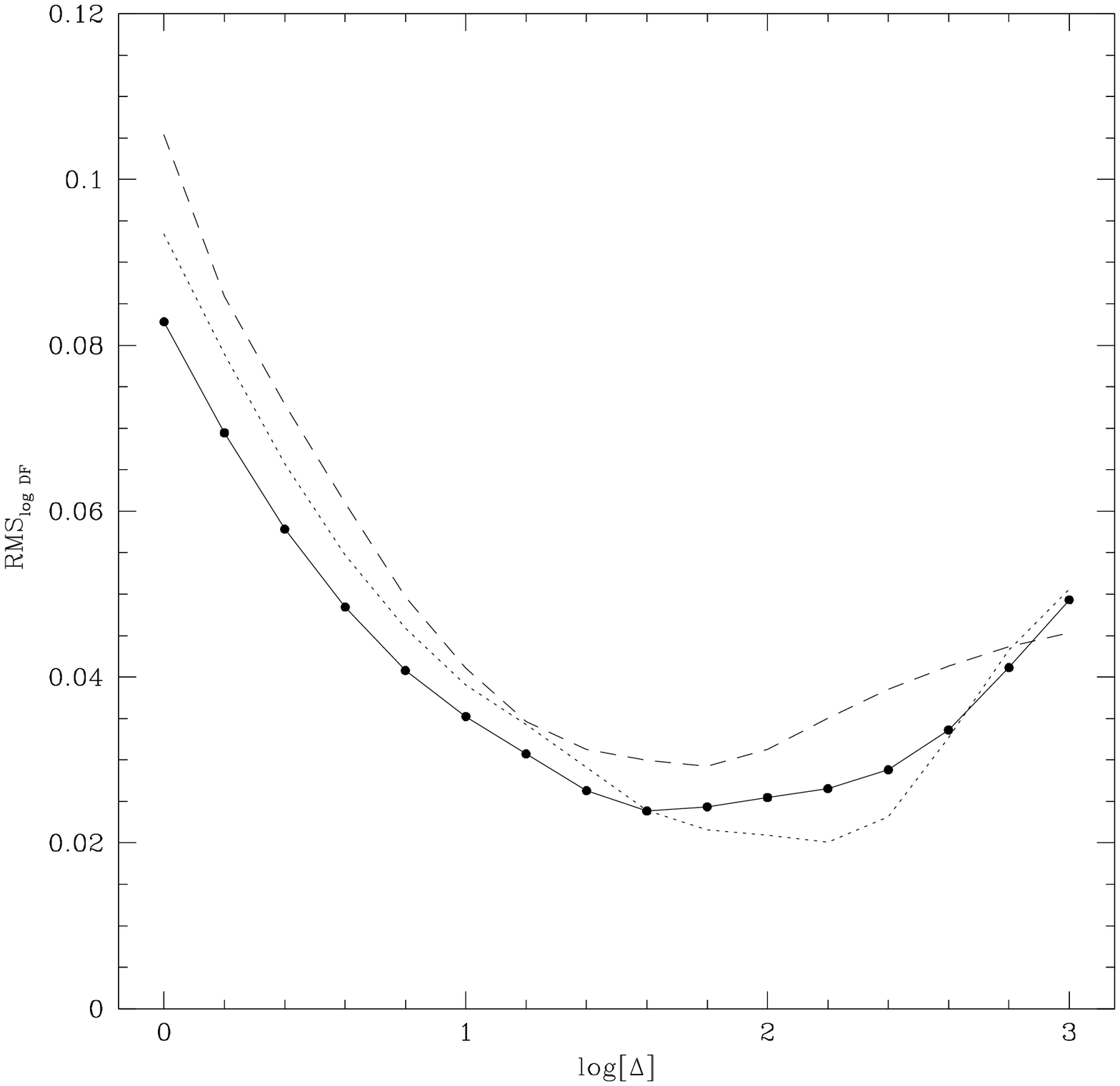}
\ifsubmode
\vskip3.0truecm
\addtocounter{figure}{1}
\centerline{Figure~\thefigure}
\else\figcaption{\figcaprmscombined}\fi
\end{figure}


\clearpage
\begin{figure} 
\epsscale{1.0}
\plotone{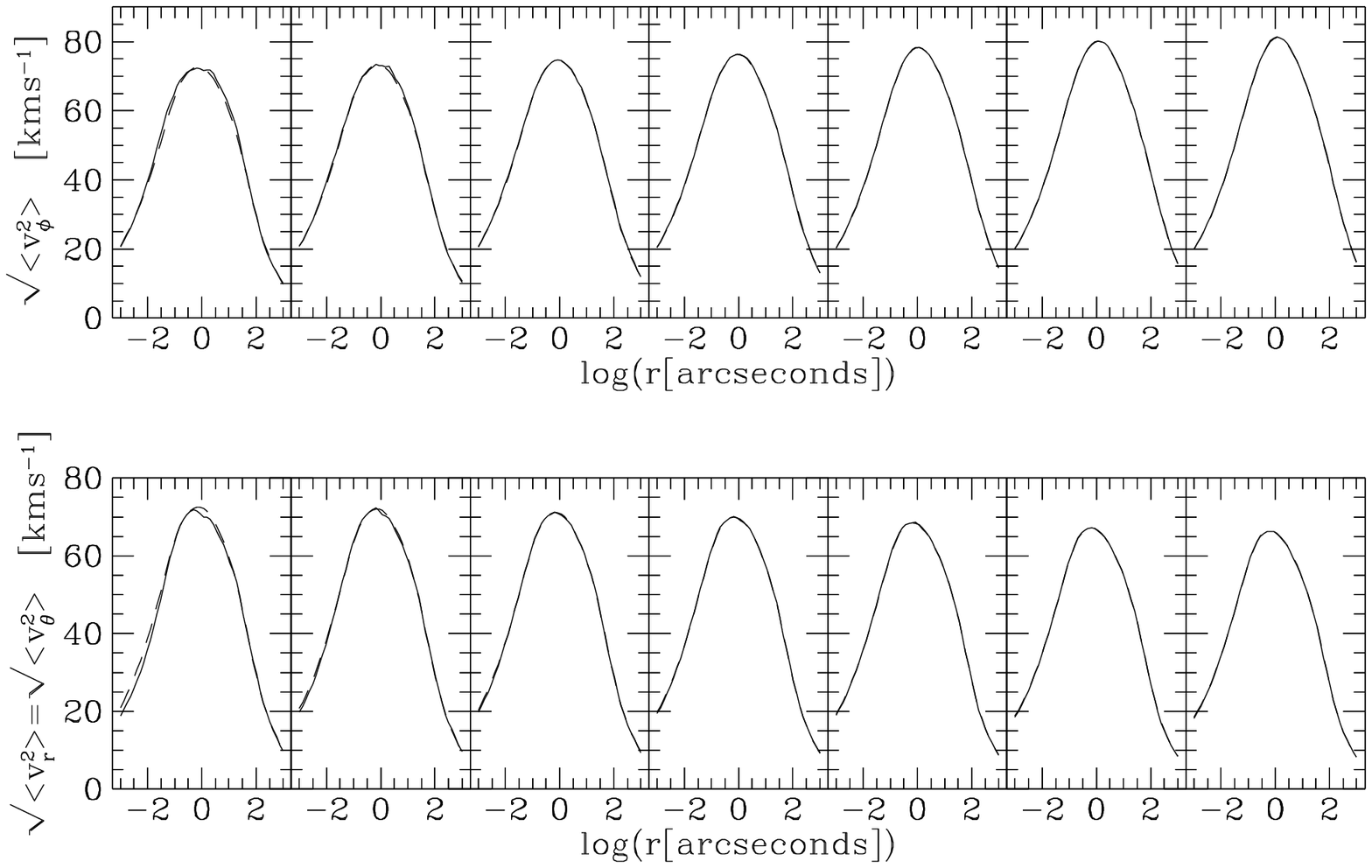}
\ifsubmode
\vskip3.0truecm
\addtocounter{figure}{1}
\centerline{Figure~\thefigure}
\else\figcaption{\figcapmomentsnoBH}\fi
\end{figure}


\clearpage
\begin{figure}
\epsscale{1.0}
\plotone{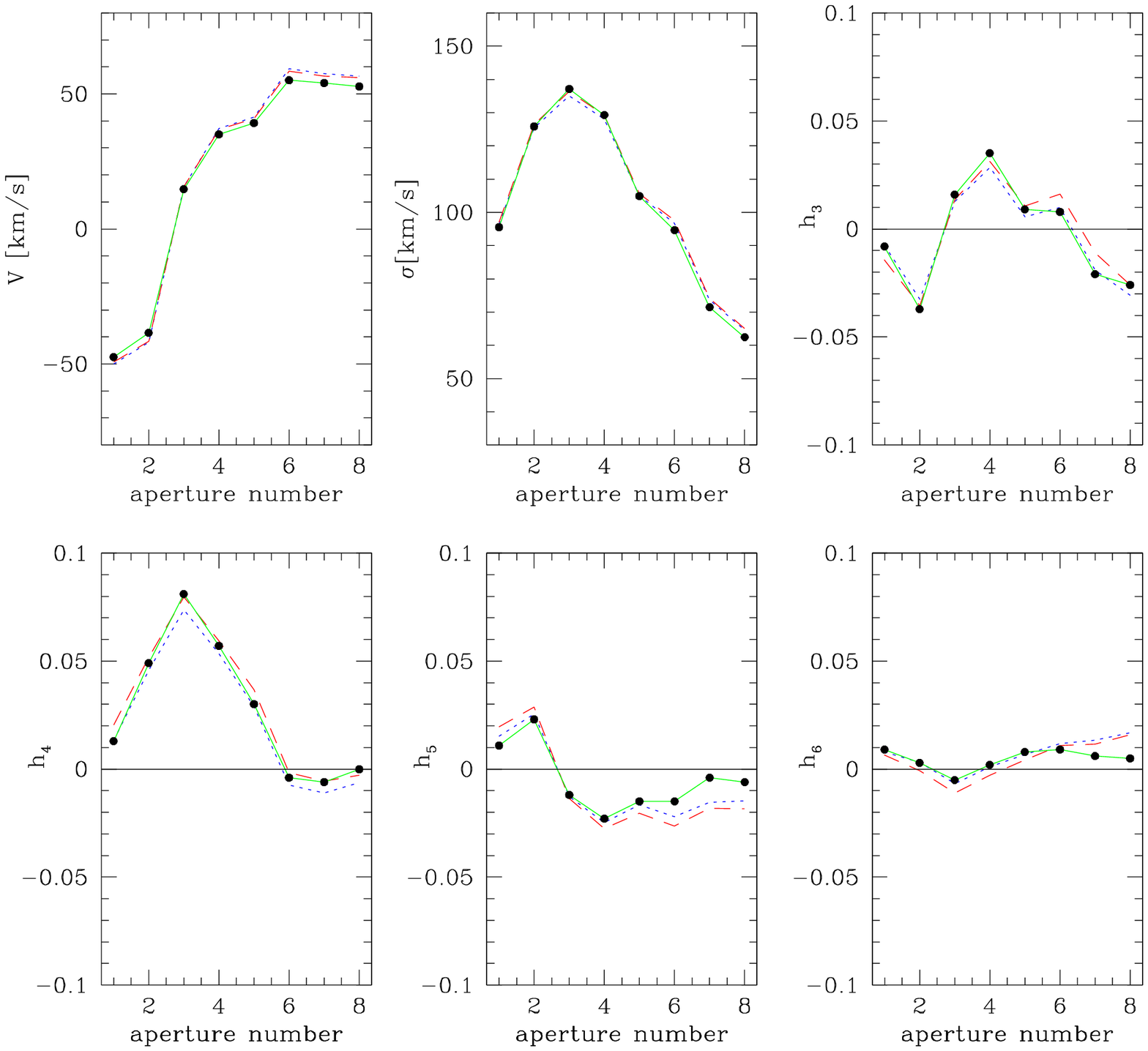}
\ifsubmode
\vskip3.0truecm
\addtocounter{figure}{1}
\centerline{Figure~\thefigure}
\else\figcaption{\figcaphstsetup}\fi
\end{figure}


\clearpage
\begin{figure}
\epsscale{1.0}
\plotone{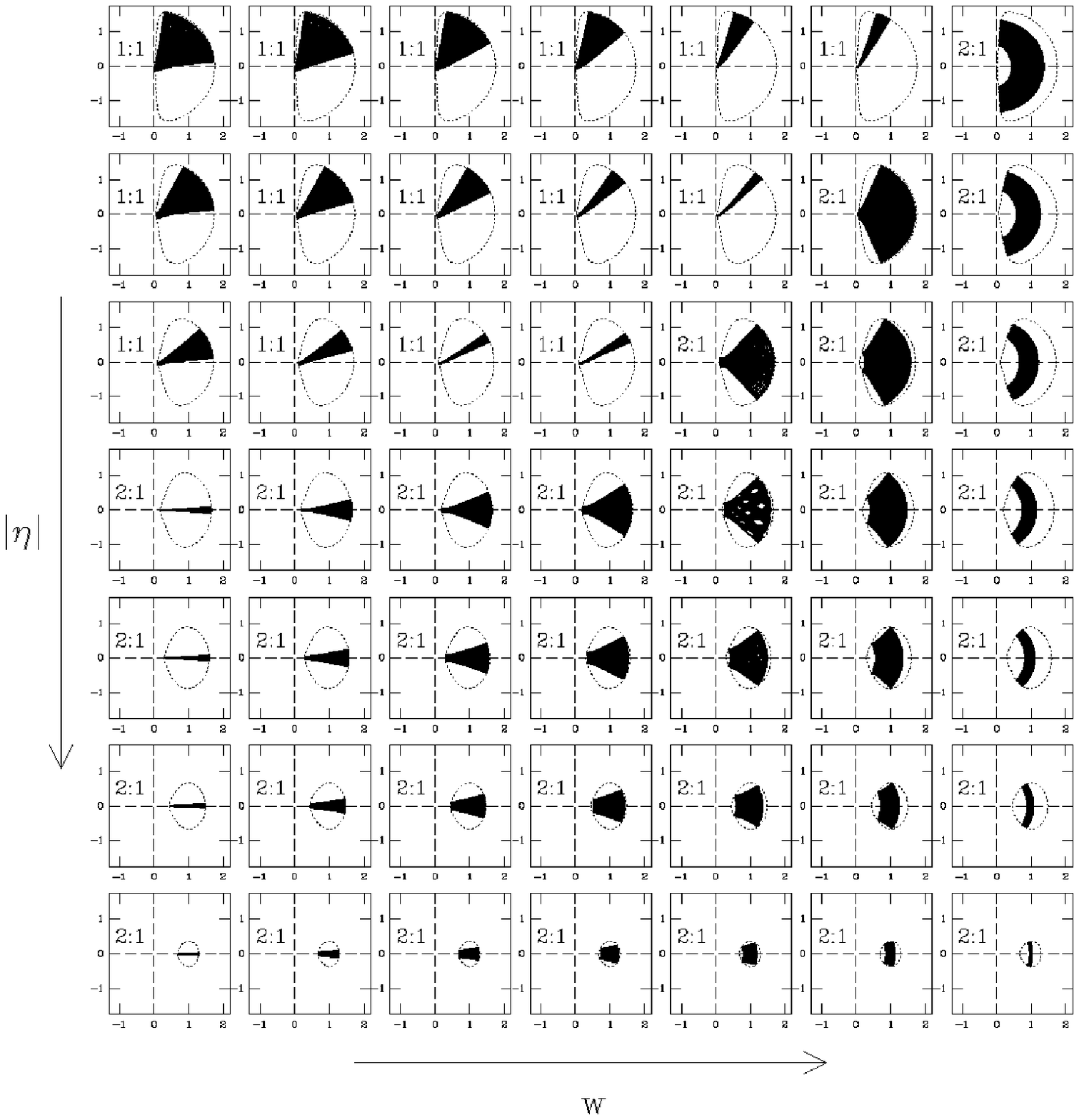}
\ifsubmode
\vskip3.0truecm
\addtocounter{figure}{1}
\centerline{Figure~\thefigure}
\else\figcaption{\figcaporbitsbh}\fi
\end{figure}


\clearpage
\begin{figure}
\epsscale{1.0}
\plotone{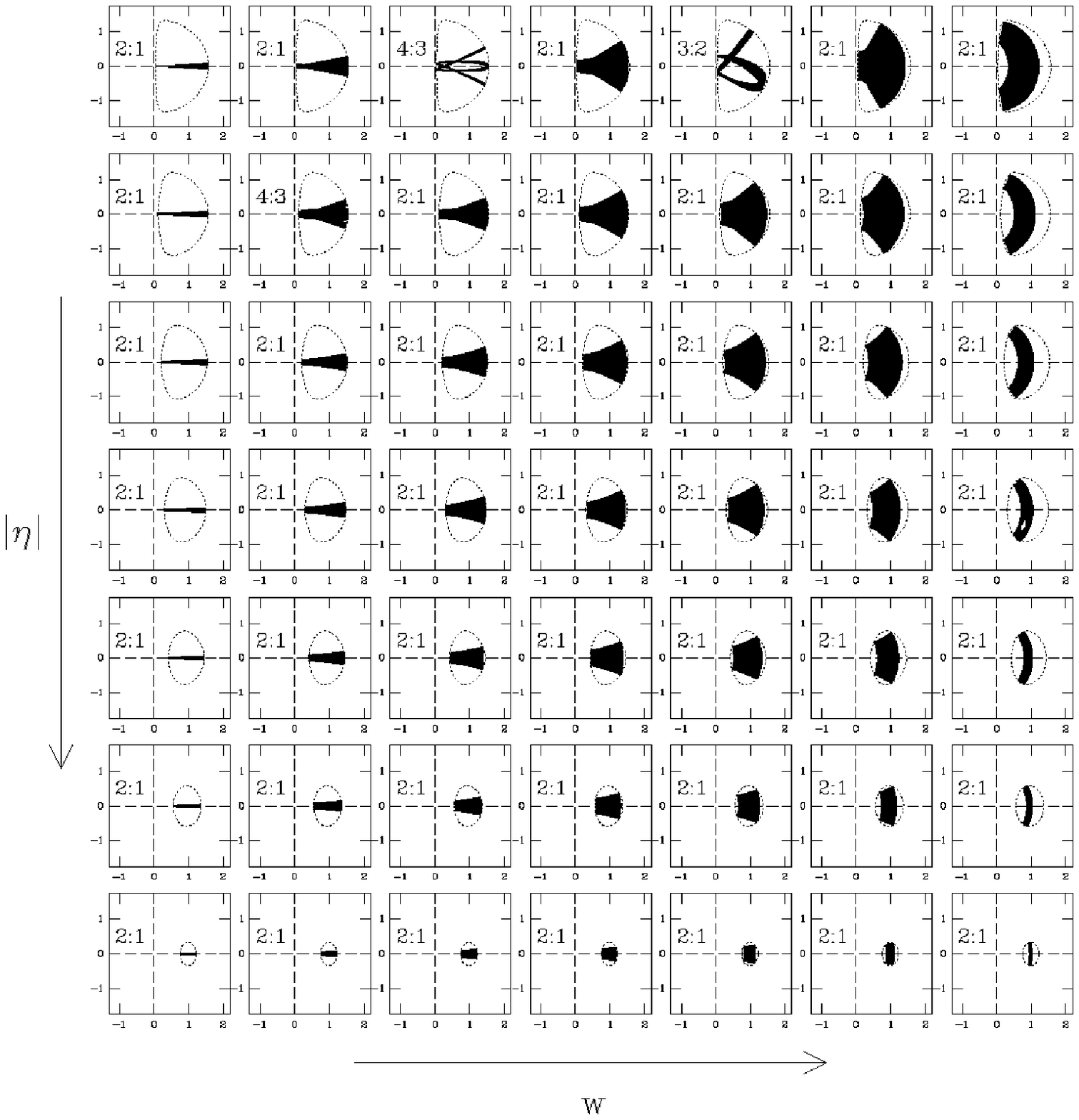}
\ifsubmode
\vskip3.0truecm
\addtocounter{figure}{1}
\centerline{Figure~\thefigure}
\else\figcaption{\figcaporbitsnobh}\fi
\end{figure}


\fi


\end{document}